\documentclass[onecolumn,amsmath,amssymb,longbibliography,notitlepage,11pt,tightenlines,superscriptaddress]{revtex4-1} %,twocolumn, pre or preprint
\usepackage{graphicx} % Include figure files
\usepackage{dcolumn} % Align table columns on decimal point
\usepackage{bm} % bold math

\usepackage{float}
\usepackage{array}

\RequirePackage{lettrine} % For dropped capitals
\newcommand{\dropcapfont}{\fontfamily{lmss}\bfseries\fontsize{26pt}{28pt}\selectfont}
\newcommand{\dropcap}[1]{\lettrine[lines=2,lraise=0.05,findent=0.1em, nindent=0em]{{\dropcapfont{#1}}}{}}

\newcommand \ellt{\ell_{{\parallel}}}
\newcommand \elln{\ell_{{\perp}}}
\newcommand \Ubend{U_{\mathrm{bend}}}
\newcommand \Ustrain{U_{\mathrm{strain}}}
\newcommand \Uwin{U_{\mathrm{subst}}}

\newcommand{\stt}{\sigma_{\theta\theta}}

\newcommand{\barm}{\bar{m}}
\newcommand{\blambda}{\bar{\lambda}}

\newcommand \q{{{\theta}}}

\newcommand \ru{{{\rm{u}}}}

\newcommand{\bfx}{{\bf x}}

\newcommand \beq{\begin{equation}}
\newcommand \eeq{\end{equation}}

\newcommand \srr{\sigma_{rr}}
\newcommand \sqq{\sigma_{\theta\theta}}

\newcommand \Keff{K_{\mathrm{eff}}}
\newcommand \Ksub{K_{\mathrm{sub}}}

\newcommand \usub{u_{\mathrm{subst}}}

\newcommand \uFvK{{u_{\rm FvK}}}
\newcommand \rmu{{\rm u}}

\newcommand \gin{\gamma_{\text{in}}}
\newcommand \gout{\gamma_{\text{out}}}
\newcommand \Rin{R_{\text{in}}}
\newcommand \Rout{R_{\text{out}}}

\newcommand \Bo{{B\!o}}
\newcommand \Boa{{B\!o}_a}
\newcommand \Bos{{B\!o_{\!a}}}
\newcommand \thetasl{{\theta_{\!a}}}
\newcommand \thetara{{\theta_{\!r}}}
\newcommand \ttau{{\tilde{\tau}}}
\newcommand{\tusub}{\tilde{u}_{\rm subst}} %{\rm win}}
\newcommand{\tubend}{\tilde{u}_{\rm bend}}
\newcommand{\tustrain}{\tilde{u}_{\rm strain}}

\newcommand \elld{\ell_{\perp}}
\newcommand \nsigma{{\hat{n}_{{\rm aux}}}}
\newcommand \naux{{\hat{n}_{{\rm aux}}}}
\newcommand \Unonlin{U_{\rm nonlin}}
\newcommand \barlambda{{\bar{\lambda}}}

\newcommand \Phiaux{{\Phi_{\tiny{\rm aux}}}}
\newcommand \sigmaaux{\tiny{\sigma_{\!\!\perp,{\rm aux}}}}

\newcommand \rs{{r_{\!a}}}
\newcommand \ms{{m_{\!a}}}

\newcommand \ellBTs{{\ell_{\rm BC}^*}}
\newcommand \ellbends{{\ell_{\rm bend}^*}}

\begin{document}

\title{Mesoscale structure of wrinkle patterns %{\emph{versus}} 
and defect-proliferated liquid crystalline phases}

% Use letters for affiliations, numbers to show equal authorship (if applicable) and to indicate the corresponding author
\author{Oleh Tovkach} %\footnote{O.T. and J.C. contributed equally to this work.}
\affiliation{Department of Physics, University of Massachusetts, Amherst, MA 01003}
\author{Junbo Chen}
\affiliation{Department of Mechanical and Aerospace Engineering, Syracuse University, Syracuse, NY 13244}
\affiliation{BioInspired Syracuse: Institute for Material and Living Systems, Syracuse University, Syracuse, NY 13244}
\author{Monica M. Ripp}
\affiliation{Department of Physics, Syracuse University, Syracuse, NY 13244}
\affiliation{BioInspired Syracuse: Institute for Material and Living Systems, Syracuse University, Syracuse, NY 13244}
\author{Teng Zhang}
\email{tzhang48@syr.edu}
\affiliation{Department of Mechanical and Aerospace Engineering, Syracuse University, Syracuse, NY 13244}
\affiliation{BioInspired Syracuse: Institute for Material and Living Systems, Syracuse University, Syracuse, NY 13244}
\author{Joseph D. Paulsen}
\email{jdpaulse@syr.edu}
\affiliation{Department of Physics, Syracuse University, Syracuse, NY 13244}
\affiliation{BioInspired Syracuse: Institute for Material and Living Systems, Syracuse University, Syracuse, NY 13244}
\author{Benny Davidovitch$^{b,}$}
\email{bdavidov@physics.umass.edu}
\affiliation{Department of Physics, University of Massachusetts, Amherst, MA 01003}

%\authorcontributions{O.T., T.Z., J.D.P, and B.D. designed research; %O.T., J.C., M.M.R., J.D.P, T.Z. and B.D. 
%all authors performed research; O.T., T.Z., J.D.P, and B.D. wrote the paper.}
%\authordeclaration{There is no conflict of interests.}

%\correspondingauthor{\textsuperscript{2}

% Keywords are not mandatory, but authors are strongly encouraged to provide them. If provided, please include two to five keywords, separated by the pipe symbol, e.g:
%\keywords{elasticity $|$ thin sheets $|$ pattern formation $|$ smectic order} 

\begin{abstract}
Thin solids often develop elastic instabilities and subsequently 
complex, multiscale deformation patterns. Revealing the organizing principles of this spatial complexity has ramifications for our understanding of morphogenetic processes in plant leaves and animal epithelia, and perhaps even the formation of human fingerprints. 
We elucidate a primary source of this morphological complexity -- 
an incompatibility between an elastically-favored ``micro-structure" of uniformly spaced wrinkles and a ``macro-structure" imparted through the wrinkle director and dictated by confinement forces. 
Our theory is borne out of experiments and simulations of floating sheets subjected to radial stretching. 
By analyzing patterns of grossly radial wrinkles we find two sharply distinct morphologies: defect-free patterns with a fixed number of wrinkles and non-uniform spacing, and patterns of uniformly spaced wrinkles separated by defect-rich buffer zones.  
We show how these morphological types reflect distinct minima of a Ginzburg-Landau functional -- a coarse-grained version of the elastic energy, which penalizes nonuniform wrinkle spacing and amplitude, as well as deviations of their actual director from the axis imposed by confinement.  
Our results extend the effective description of wrinkle patterns as liquid crystals (H. Aharoni {\emph{et al.}}, Nat. Commun. 8:15809, 2017), and we highlight a fascinating analogy between the geometry-energy interplay that underlies the proliferation of defects in the mechanical equilibrium of confined sheets and in thermodynamic phases of superconductors and chiral liquid crystals. 
\end{abstract}

\maketitle
%\thispagestyle{firststyle}
%\ifthenelse{\boolean{shortarticle}}{\ifthenelse{\boolean{singlecolumn}}{\abscontentformatted}{\abscontent}}{}

% If your first paragraph (i.e. with the \dropcap) contains a list environment (quote, quotation, theorem, definition, enumerate, itemize...), the line after the list may have some extra indentation. If this is the case, add \parshape=0 to the end of the list environment.

\begin{figure}
%\vspace{-2cm}
\centering 
\begin{center} 
\includegraphics[width=0.5\textwidth]{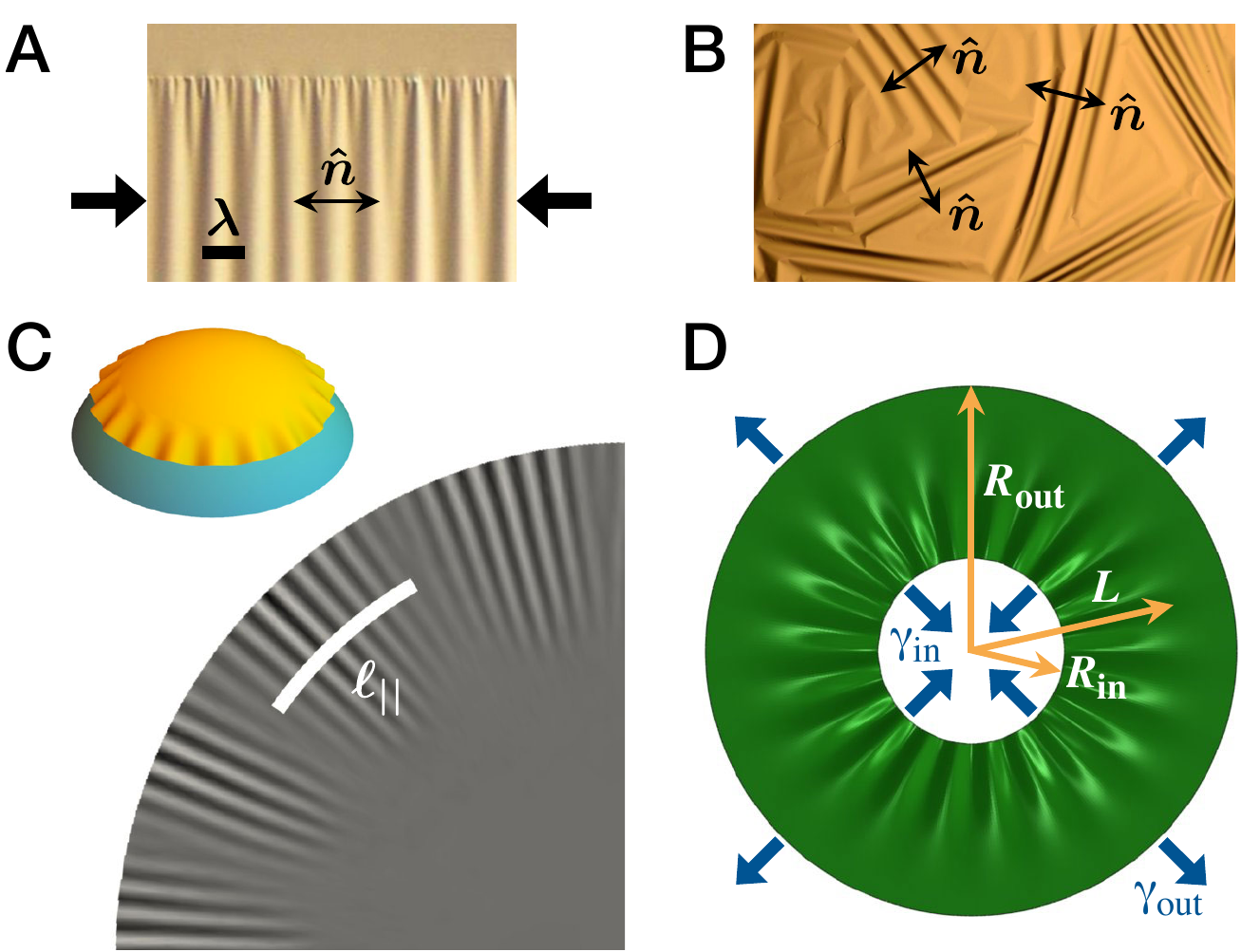} 
\end{center}
\caption{
{Director and wavelength of wrinkle patterns.} 
{(A)} Uniaxial confinement of an ultrathin sheet floating on water. The sheet undulates along the director $\hat{n}$ to suppress compression in this axis, forming a parallel array of wrinkles with a wavelength, $\lambda = 2\pi (B/\rho g)^{1/4}$ (black bar). Deviations from this pattern appear only at the edge, where a liquid meniscus affects a wrinkling cascade [from Ref.~\cite{Huang10}]. (B) 
Wrinkle pattern within an ultrathin spherical shell floating on a liquid bath, 
where the director varies across domains of parallel, uniformly-spaced wrinkles.
%which exhibits a ``domain wall'' wherein the wrinkle director undergoes a sharp change; this structure may be described by analogy to the elasticity of smectic phases \cite{Aharoni17}. 
%The director imposed on a %an (infinitely thin) polygonal patch of an ultrathin spherical shell by floating it on a liquid bath, consists of a ``domain wall'' is discontinuity, yielding a ``domain wall'' in a physical (ultrathin) shell, whose width may be described by analogy to the elasticity of smectic phases \cite{Aharoni17}. 
(C) Simulation of a circular thin sheet attached to a ball of stiff springs of constant $\propto \Ksub$, where confinement imposes a purely-bent director ($\hat{n} = \theta$), incompatible with wrinkles separated by a constant, locally-favored wavelength, $\barlambda = 2\pi (B/\Ksub)^{1/4}$ \cite{Hohlfeld15,Davidovitch19}. %{\color{blue} TFT determines the radial extent and amplitude of wrinkles, as well as the stress field, through expressions analogous to Eqs.~(\ref{eq:TFT-1},\ref{eq:slaving}), see Eqs.~38, 41-43 in Ref. \cite{Hohlfeld15}.} 
The pattern exhibits defect-rich buffers and pronounced amplitude modulations, characterized by a length $\ellt$ (white bar shows the prediction of our theory,  Eq.~\ref{eq:ell-amp-1}).
(D) The Lam\'e set-up is an annular sheet under coaxial, co-planar tensile loads, $\gin,\gout$. 
%(e) If the ratio $\tau = \gin/\gout$ exceeds a threshold, radial wrinkles emerge around the inner edge to suppress azimuthal compression. When the sheet is floated on a liquid, and a Bond number $\Bo = \rho_{liq} g\Rin^2/\gin$ is sufficiently large the characteristic pattern resembles panel c, with defect-rich buffers and pronounced amplitude modulations between them (from simulation).     
}   
%\vspace{-0.1cm}
\label{fig:1}
\end{figure}

\dropcap{T}hin solid bodies 
%sheets and shells 
tend to suppress compression by developing wrinkles -- elongated periodic undulations.
%with small ``wavelength''.  
Wrinkle patterns are ubiquitous due to the broad range of conditions that generate compression: boundary loads \cite{Cerda03}, incompatible topographical constraints \cite{Hure12,Stoop15,Paulsen19}, differential swelling \cite{Klein11}, %,Gemmer16}, %thermal/chemical 
expansion on soft substrates \cite{Bowden98,Breid11}, and growth in confined spaces \cite{MahaVili,Kucken04}, have all been recognized as potential drivers of wrinkled morphologies.
%Notwithstanding  the beautiful example that wrinkle patterns provide for the spontaneous emergence of a complex morphology through a sequence of  symmetry-breaking instabilities, %under featureless forces or geometric constraints, 
%Furthermore, wrinkling mechanics can be harnessed for non-invasive reading of the principal stress directions of human skin by a surgeon preparing to make a cut, to determine the stiffness of a subphase \cite{VellaLit}, or to measure the thickness of ultrathin sheets \cite{Huang07,Schroll13,Stafford07}.  
A basic picture, often used to model %wrinkling 
these phenomena, is uniformly-spaced undulations along parallel lines (lower part of Fig.~\ref{fig:1}a). However, most observed patterns differ significantly from such a simplistic picture, demonstrating instead how multi-scale patterns emerge under smooth, featureless forcing.

%. However, while 
%this ``monochromatic'' pattern may emerge only under uniaxial confinement of a uniform sheet attached to a uniform substrate. Such an ideal picture may be useful for analyzing %is often foassumed when analyzing small portions of a wrinkled sheet, but mischaracterizes   
%the complexities encountered in most observed patterns which emerge under non-uniform, non-uniaxial confinement conditions. 

A predominant source of complexity here is a %the 
conflict between two primary features: a {\emph{wavelength}} $\lambda$ ({\emph{i.e.}} distance between nearby peaks), and a {\emph{director}} $\hat{n}$ -- %, which dictates 
the axis along which the sheet undulates. 
%undulatory axis. 
The former is a {\emph{micro-scale}} object, typically determined by a local balance of bending rigidity and the stiffness of an ``effective substrate'' \cite{Cerda03,Paulsen16},  %(namely, $\lambda$ vanishes with the thickness of the confined solid), %decreases indefinitely as the thickness of the confined body is reduced), 
whereas the latter is a {\emph{macro-scale}} field that reflects the confining topography and lateral forces exerted on the body 
%variation is at a {\emph{macro-scale}} (namely, 
($|\nabla \hat{n}| \ll \lambda^{-1}$) \cite{MansfieldBook,Stein61,Pipkin86}. 
The basic pattern of perfectly parallel wrinkles %, namely a constant director $\hat{n}$ 
%(Fig.~\ref{fig:1}a), 
emerges when a sheet is confined uniaxially. However, deviations from this ideal picture occur when either the director or the locally-favored wavelength are non-uniform (Fig.~\ref{fig:1}a-b).     
Here we address a %different type of 
{\emph{mesoscale}} structure -- namely, at scales intermediate between $\lambda$ and $|\nabla \hat{n}|^{-1}$ --      
 that emerges when relaxation of compression implies a {\emph{bent}} wrinkle director, underlying radially-oriented wrinkles (namely, $\hat{n} \approx \hat{\theta}$). 
%In such situations (Fig.?), relaxation of compression requires radial wrinkles, however, 
% (namely, 
%undulations along an azimuthal axis 
%$\hat{n} \approx \hat{\theta}$). 
%Since a
%A ``monochromatic'' pattern of $m_0$ radial wrinkles requires spacing that varies with radial distance, $\lambda(r) \approx (2\pi/m_0) r$, which is
%it is clearly 
In a ``monochromatic'' pattern of $m_0$ wrinkles the spacing varies with radial distance, $\lambda(r) \approx (2\pi/m_0) r$, and is thus  
incompatible with any locally-favored $\lambda$ that is not %proportional to 
$\propto r$. 
We discover that prominent  
%mesoscale motif that emerges from 
%mesoscale 
outcomes of this conflict %comprises 
are modulations of the wrinkle amplitude over a mesoscale scale $\elln$,
which depends on the locally-favored wavelength %$\lambda$ 
and confinement conditions, and the 
%physical parameters given by the confinement conditions, and the 
%\gg \lambda$, and %that enable 
proliferation of defect-rich regions, where the number of wrinkles varies sharply (Figs.~\ref{fig:1}c,\ref{fig:2}b,\ref{fig:2}d). 
%A central result of our work is an invariant formula (Eq.~\ref{eq:ell-amp-1}) 
%%%that predicts the characteristic scale over which the amplitude varies, 
%for the amplitude variation length $\elln$, and consequently the %the %(azimuthal)  
%mesoscale extent of defect-free wrinkled zones, in terms of the locally-favored wavelength %$\lambda$ 
%and physical parameters given by the confinement conditions. 
%This prediction is compared with experimental observations and numerical simulations. 
The combined effect %crucial %essential 
%role 
of defects and %strong 
amplitude modulations 
%in shaping the pattern  
 %interplay between the proliferation of defects and strong amplitude modulations
%The emergence of strong amplitude modulations 
suggests an analogy between mesoscale wrinkling phenomenology and   
%that an effective theory of mesoscale wrinkling phenomenology resembles 
%phenomenology exhibited by wrinkle patterns 
%is not given by a mere analogy to 
%is not merely analogous to the 
%is more complex than a simple 
%smectic phase \cite{Aharoni17}, indicating similarity 
%a broad class of 
defect-proliferated phases of %states %in condensed matter, found {\emph{e.g.}} 
%in 
liquid crystals and superconductors   
%liquid-crystalline or superconducting states of matter 
\cite{DeGennesBook,Renn88}. 
%as was suggested recently \cite{Aharoni17}.    

%%%%%%%%%%%%%%%%%%%%
%\vspace{-2cm}
%%%% FIGURE 1 %%%%%%%%%%%%%%%%

%Furthermore, for a given confinement, we characterize a morphological transition between such a defect-proliferated              

%undergoes sharp variations.   
\section*{Model system}

A bent director %that dictates radially-oriented wrinkles 
may appear whenever a confined solid with a ``target'' ({\emph{i.e.}} metrically-favored)  
 %whose natural (``target'') metric is characterized by a 
 Gaussian curvature $G_{\rm tar}$ is forced to reside near a ``substrate'' whose shape has a curvature $G_{\rm sub} > G_{\rm tar}$ ({\emph{e.g.}} %stamping 
 a naturally-planar sheet attached to %onto 
 a liquid drop \cite{King12} or a rigid sphere \cite{Hure12,%Grason13,Azadi14,
 Davidovitch19}, Fig.~{\ref{fig:1}c). However, for concreteness and clarity we focus %will focus here 
 on the classical Lam\'e set-up, whose numerous variants were studied extensively, %in recent years, 
 allowing us to exploit a host of experimental \cite{Huang07,Toga13,Pineirua13}, numerical \cite{Taylor15}, and analytic techniques \cite{Davidovitch11,Taylor15}. %, and theoretical results \cite{Davidovitch11,Taylor15}.
 %, for addressing the emerging mesoscale structure. 
 A schematic of the model 
 is shown in Fig.~\ref{fig:1}d:  
 %-- a thin annulus, whose inner edge ($r=\Rin$) is pulled inward. Figure ? shows schematically the Lam\'e set-up: 
a thin solid annulus of thickness $t$ and radii $\Rin \ll \Rout$ is attached to a ``Winkler substrate'' of stiffness $\Ksub$ ({\emph{e.g.}} $\Ksub = \rho g$ for a liquid bath of density $\rho$), and subjected to tensile loads, $\gin, \gout$ that pull the edges inward and outward, respectively. The stretching and bending moduli of the sheet are $Y = Et \gg \gin,\gout$, and $B = Et^3/12(1-\Lambda^2)$, where $E,\Lambda$ are the Young's modulus and Poisson's ratio, % of the solid, 
respectively. The problem is governed by three dimensionless groups, to which we refer, respectively, as {the} ``confinement'', % ratio, 
tensional ``bendability'', and a {\emph{Bond}}-like parameter ({\emph{i.e.}} %characeristic 
the ratio between {substrate stiffness and tensile stress}):   
\begin{equation}
\tau  = \frac{\gin}{\gout} \ \  ;\  \ \epsilon^{-1} = \frac{\gin \Rin^2}{B} \  \ ; \  \ 
%\Bo = \frac{\gin}{\Ksub\Rin^2}  \ , 
\Bo = \frac{\Ksub\Rin^2}{\gin}  \ . 
\label{eq:nondim-param}
\end{equation}    
%\jp{We focus on thin films that have high bendability ($\epsilon^{-1}\gg 1$).}
{We study thin, highly bendable sheets ($10^4<\epsilon^{-1}<10^8$).} 
%($\epsilon^{-1}\gg 1$).}
%If the sheet is highly bendable ($\epsilon^{-1}\gg 1$) and 
If $\tau$ 
%sufficiently thin and $\tau$ 
%the ``confinement ratio'' $\tau = \gin/\gout$ 
exceeds a finite threshold ($\gtrsim 2$),
hoop confinement emerges in  
an annulus, %zone 
$\Rin<r<L$, %is subjected to hoop confinement, 
which expands upon increasing $\tau$, 
% In the high-bendability regime, $\epsilon^{-1}\gg 1$, 
%and hoop 
and compression is suppressed through azimuthal undulations, yielding radially-oriented wrinkles. 
Macro-scale features of the pattern are governed by $\tau$ (see below), 
%specifically the radial extent $L$ and the radial stress profile, see explicit expressions in \cite{SI}), 
whereas the wavelength scales as $\lambda(r) \sim \epsilon^{1/4}$
%, and vanishes in % such that it vanishes in 
%the singular limit $\epsilon \to 0$ 
\cite{Huang07,Davidovitch11,Pineirua13,Paulsen16}. 
Experiments exhibit a largely uniform wavelength when the substrate stiffness is strong ($\Bo \gg 1$) \cite{Pineirua13},  
%, experiments exhibit 
%a largely uniform wavelength \cite{Peinerua13},  
%that explored 
%reported the characteristic wavelength at 
%various ranges of the substrate-dominated ($Bo \ll 1$) and tension-dominated ($Bo \gg 1$) regimes, 
%indicated a laregly uniform wavelength when the substrate stiffness is strong,   
%($\lambda(r) \approx 2\pi (B/\Ksub)^{1/4}$) 
%in the substrate-dominated regime, 
and a constant wrinkle number ({\emph{i.e.}} $\lambda(r) \propto r$) for a weak substrate \cite{Huang07,Toga13,Jooyan18},  
suggesting that: %the form:  
\begin{gather}
\lambda(r)/\Rin \approx \epsilon^{1/4} \cdot %F(r; \tau,\Bo)  \nonumber \\
%F(r; \tau,\Bo) = 
\ \left\{ \begin{array}{ll}
         2\pi  \cdot {B\!o}^{-1/4}  & \ \  \mbox{if $\Bo \gg  1$} \\
        C(\tau) \cdot r/\Rin &   \ \ \mbox{if $\Bo \ll  1$} \ , \end{array} \right.  
\label{eq:lambda-1}
\end{gather} 
(where $C(\tau)$ is some smooth function).
%with a prefactor $F(r,Bond)$, 
%such that it vanishes in the singular limit $\epsilon \to 0$ \cite{Huang07,Davidovitch11,Peinerua13,Paulsen16}. The relative effect of substrate stiffness and radial tension on the micro-scale $\lambda(r)$, encapsulated in the function $F(r; \tau, Bo)$, and the consequent mesoscale structure, are the primary focus of our study.   
%and is in a manner that depends on the {\emph{Bond}} parameter.    

 %are governed primarily by the ratio $\tau$, whereas the locally-favored value of $\lambda$ determined    
%Various variants of this model system were studied extensively in recent years, allowing us to exploit a host of experimental methods \cite{Huang07,Toga13, Peinerua13,Stone16}, numerical techniques \cite{Taylor15}, and theoretical results \cite{Davidovitch11,Davidovitch12}, for addressing the emerging mesoscale structure.  
%and a couple of typical patterns observed in experiment and simulation. 
%%%%%%%%%%%%%%%%%%%%%%%%%%%%%%

%%%%%%%%%%%%%%%%%%%%
%%%% FIGURE 2 %%%%%%%%%%%%%%%%
\begin{figure*}
\centering 
\begin{center} 
\includegraphics[width=1.0\textwidth]{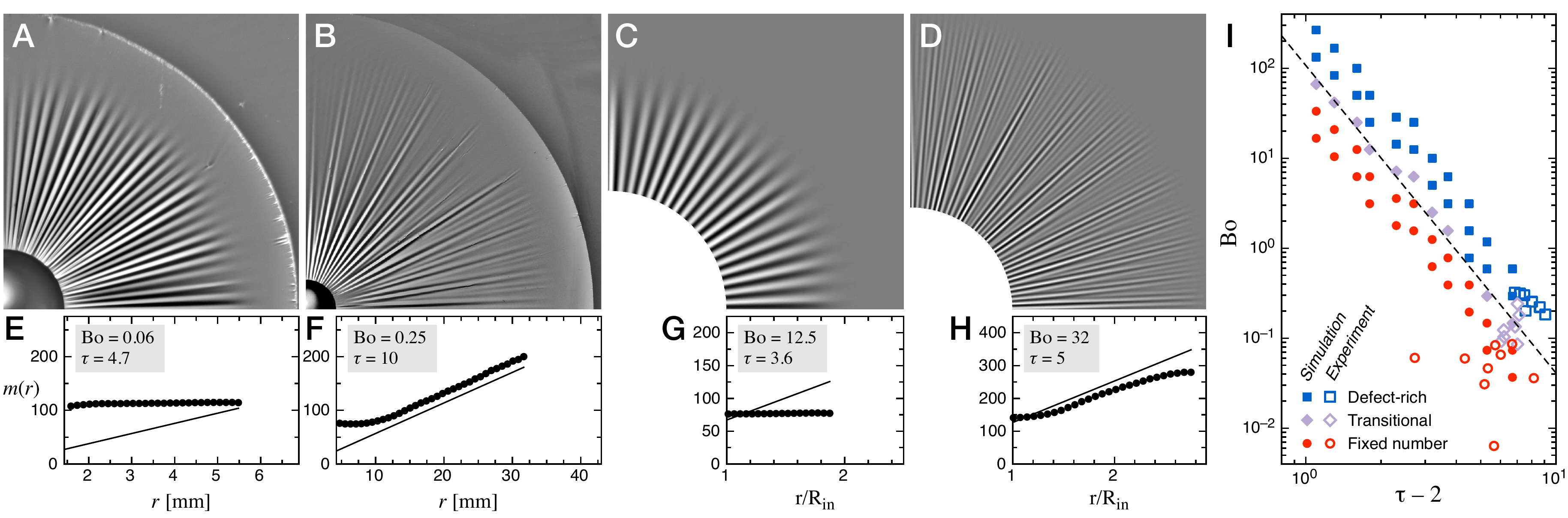} 
\end{center}
\caption{
Characteristic patterns observed in experiment (A-B) and simulation (C-D) for various values of the control parameters %$\tau$ and $\Bo$ 
(Eq.~\ref{eq:nondim-param}), as listed in the figure. 
(In panels A,B, we subtracted a background obtained by a uniform Gaussian filter.)
{In the experiment, the Bond number is controlled primarily by the droplet size, and we use larger sheets to accommodate longer wrinkles. 
(A) $\Rin=1.4$ mm, $\Rout=6.9$ mm. 
(B) $\Rin=4.3$ mm, $\Rout=43$ mm.} 
%(A): $\Rin=1.4$ mm, $\Rout=6.9$ mm, $t=62$ nm.   
%$\Bo=0.06$, $\tau=4.7$. %$\Rin=1.4$ mm, 
%(B): $\Rin=4.3$ mm, $\Rout=43$ mm, $t=308$ nm.   %$\Bo=0.25$, $\tau=10$. %$\Rin=4.3$ mm, 
%(C,G): $\Bo=12.5$, $\tau=3.6$. 
%(D,H): $\Bo=32$, $\tau=5$. 
{Panels (E-H) show the wrinkle number $m$ versus radial coordinate $r$
%number of undulations at a radial distance $r$, 
for the images above them.} 
%Representative images in experiments (b) and simulations (c) with a fixed number of radial wrinkles are observed for large values of $Bo$, where the substrate effect is negligible in comparison to (radial) tension. (d-e) Representative images in experiments (d) and simulations (e) in which the wrinkle wavelength is approximately constant are observed for small values of $Bo$, where the substrate effect sufficiently strong to determine a locally-favored wavelength. In such patterns we observe defect-rich zones, and pronounced modulations of the wrinkle amplitude.} 
{Solid lines show the value that would result from a local balance of substrate stiffness and bending: $\bar{m}(r) =2\pi r/\barlambda= r(\rho g/B)^{1/4}$.} 
Note that the boundary condition at $r=\Rin$ tends to increase the wrinkle number in its vicinity.
(I) A ``phase diagram'', spanned by the parameters $\Bo$ and $\tau$ %(and small values of $\epsilon$, such that $\epsilon/\Bo =Cst$), 
is divided into two regimes where wrinkle patterns are defect-rich (blue symbols, where the undulation wavelength $\approx \barlambda$, panels B,D and F,H), and defect-free (red symbols, where the wrinkle number is constant, panels A,C and E,G). 
%The solid (dashed) curves show best-fit of simulation (experimental) data to the theoretical prediction (\ref{eq:Boc}). 
Dashed line: %shows 
Empirical %{theoretically-predicted 
scaling relation, Eq.~(\ref{eq:Boc}), with a fitted numerical prefactor.} 
%, where the unknown numerical prefactor has been fit to the data.} 
%The difference in the numerical prefactors between the two curves is attributed to the distinct ranges of the bendability parameter $\epsilon^{-1}$, Eq.~(\ref{eq:nondim-param}), explored in experiments and simulations (see {\emph{SI}}).}
%exhibits twoqualitatively-distinct types of patterns in parameter regimes that are separated by a curve $\Bo_c(\tau)$.  
%, which may depend on the fixed ratio $\epsilon/\Bo$. Above this curve we find in experiments and simulations patterns with a fixed number of radial wrinkles; below it we find defects-proliferated patterns with a fixed wavelength (well described by Eq.~\ref{eq:lambda-1}), that contains defect-rich zones separated by defect-free zones with strong amplitude modulations
\label{fig:2}
\end{figure*}

Our experiments and simulations explore a broad range of $\Bo$ values.
%, with \jp{$10^4 < \epsilon^{-1} < 10^8$.} % in experiment and $10^{-8} < \epsilon < 10^{-5}$ in simulation. 
The simulations employ a finite element method, similarly to \cite{Taylor15}, with 
%the important exception that we use 
a (locally-hexagonal) disordered mesh to suppress spurious lattice effects on the pattern. 
%BD (5-9-19): confusing here. Need to mention it later on. 
%We fix the softness parameter, $\sqrt{\rho g B}/\gamma = 0.01$ in the simulations. 
The experiments use spin-coated, ultrathin polystyrene sheets (thickness $60<t<308$ nm) of circular shape (radius $6.9<\Rout<44$ mm) that are floated on a liquid bath and subjected to radial tension $\gout = \gamma_{lv} = 70 \pm 2$ mN/m at $r=\Rout$. 
A liquid drop placed at the center of the sheet forms a circular contact line of radius $\Rin$, yielding capillary-induced tension, $\gin > \gamma_{lv}$ \cite{Schroll13}; see {\emph{Supplementary Information (SI)}}. 
%that pulls the sheet inward at the contact line ($r=a$). 
%The value of $\gamma_i$ is given by formula that reflects a balance of the stretching modulus of the sheet ($Y = Et$, where $E$     
 
The images in Fig.~\ref{fig:2} show two contrasting %morphological 
responses: 
a spatially-constant wrinkle number (\emph{i.e.} non-constant wavelength, $\lambda \propto r$), and patterns with nearly-constant wavelength ($\lambda \approx \lambda_0$ so that $m(r) \approx 2\pi r/\lambda_0$, as shown in Figs.~\ref{fig:2}b,\ref{fig:2}d with measurements of $m(r)$ in Figs.~\ref{fig:2}f,\ref{fig:2}h). 
The latter patterns are significantly more complex, as they require the proliferation of defects, \emph{i.e.}, points where wrinkles are created away from boundaries. %inside the wrinkled zone ($\Rin<r<L$).  
These patterns also exhibit strong modulations of the wrinkle amplitude, such that defect-rich regions occur at amplitude-suppressed zones.
Figure~\ref{fig:2}i collects our observations %from simulations and experiments 
into a phase diagram. 
The observed transition between defect-free and defect-rich states forms a curve $\Bo_c(\tau)$ in the parameter plane $(\tau,Bo)$, 
which we find to scale as: 
%inversely with the cubic power of the radial width of the wrinkled zone:   
\begin{equation}
\Bo_c (\tau) \sim (\tau -2)^{-3.4 \pm 0.3} \ . 
\label{eq:Boc}
\end{equation}

\section*{Theory}
The above findings motivate us to focus %our study 
on 
%{study} 
the {morphologically-rich} regime {at large Bond number}. This section and a subsequent one describe succinctly our theoretical approach, delegating many technical details to {\emph{SI}}.    
%where multi-scale perturbation theory can be employed. 
%The starting point of our analysis 
Our starting point is %Let us first recall  
``tension field theory'' (TFT), which provides the leading-order {F\"oppl-von K\'arm\'an} (FvK) elastic energy 
%equations of mechanical equilibrium 
at the singular limit of infinite bendability \cite{Davidovitch11}. 
%{\bf Mention here that TFT is the basis for FT analysis, whose validity (for $\Bo \gg 1$) requires $\epsilon \Bo \ll 1$, and refer to SI.}
Positing that the ratio between the compressive (hoop) and tensile (radial) components of the stress tensor must vanish as $\epsilon \to 0$, in-plane force balance in the wrinkled zone ($\Rin<r<L$), and matching to the purely-tensile (unwrinkled) region yield: %the radial stress $\srr(r)$ and the extent  
 \begin{gather}
(\Rin\!<\!r\!<\!L): \  \srr \approx \gin \Rin/r  \ \ ; \ \  L \approx  \Rin \tau/2 \ . 
\label{eq:TFT-1}
 \end{gather} 
%implies that $\srr(r,\theta) \approx \gin \Rin/r$ in the wrinkled zone, 
%$\Rin<r<L$, 
%and matching to the purely-tensile (unwrinkled) region %($L<r<\Rout$) 
%shows that the extent of the wrinkled annulus is $L \approx \tfrac{1}{2} \Rin \tau$. 
Suppression of hoop compression 
requires matching the contraction, ${\rm u}_r/r$, due to the displacement ${\rm u}_r(r)$ underlying
(\ref{eq:TFT-1}), to  
%implies that 
the fraction of latitudinal length $\Phi^2$  ``wasted'' by wrinkles:%, yielding:}  
%to the contraction radial displacement underlying   :
%requires the fractional excess latitudinal length $\Phi^2(r)$ ``wasted'' 
%by wrinkles %per wrinkle 
%in a hoop of radius $r$ 
%to be: 
 \begin{gather}
%\tfrac{1}{2\pi} \!\! 
\Phi^2(r) \equiv  
\frac{1}{2\pi}\int_0^{2\pi} \!\!\!\!\!\!d\theta \left(\sqrt{1+ (\tfrac{1}{r}\tfrac{\partial\zeta}{\partial \theta})^2} \ - 1\right)
 \ \approx \   \frac{1}{4\pi r^2} \int_0^{2\pi} \!\!\!\!\!\!d\theta (\tfrac{\partial\zeta}{\partial \theta})^2    \nonumber \\
 =  2 (\gout/Y) {(L/r) \log(L/r)} \ , 
 \label{eq:slaving}
\end{gather}
where $\zeta(r,\theta)$ is the deflection from the plane \cite{Davidovitch11}.    
%$\zeta(r,\theta)$ is the deflections 
Notably, these macro-scale features characterize the singular limit $\epsilon \to 0$, being indifferent to the wavelength $\lambda(r)$. 
The corresponding limit value of elastic energy (in comparison to a state with uniform strain $\gout/Y$) 
does not depend on $\epsilon$ or $\Bo$, namely:
%depends only on $\tau$:} 
\begin{equation}
U_{\rm dom}(\tau) \approx 
- \frac{\pi \Rin^2\gout^2}{Y} \tau^2 ( \log{\frac{\tau}{2}} + \Lambda - \frac{1}{2})   \ . 
\label{eq:Udom}
\end{equation}  
In order to remove this shape degeneracy,  %the degeneracy left over by the TFT solution,  
%determine $\lambda(r)$, 
one must minimize a ($\epsilon$-dependent) contribution to the FvK energy that penalizes bending and substrate deformation, by solving the 
%corresponding Euler-Lagrange (equivalently, $1^\text{st}$ FvK) 
$1^\text{st}$ FvK equation (assuming $|\partial_\theta\zeta| \gg r|\partial_r\zeta|$): 
\begin{gather}
%(a) \ 
%\left({\cal L}_0  + {\cal L}_1 \right) \Psi(r) = 0  \label{eq:Fvk-1} \\
({\cal L}_0 +  {\cal L}_1) \zeta(r,\theta) = 0    \label{eq:FvK-1} \\
%{\cal L}_0\Psi(r) - \srr(r) \Psi''(r)  = 0  
%\    ;  \  
%(b )\ \ 
%\text{where:} \ 
%{\cal L}_0  \!=\! B \left(\tfrac{m_0}{r}\right)^4 \!-\! \sqq\left(\tfrac{m_0}{r}\right)^2 \!+\! \Ksub   \nonumber \ ; \ 
%{\cal L}_1 =   - \srr \tfrac{d^2}{dr^2} \ , 
\text{where:} \ {\cal L}_0  \!=\! B \tfrac{1}{r^4}\tfrac{\partial^4}{\partial\theta^4} \!-\! \sqq  \tfrac{1}{r^2}\tfrac{\partial^2}{\partial\theta^2}
\!+\! \Ksub   \nonumber \ ; \ 
{\cal L}_1 =   - \srr \tfrac{\partial^2}{\partial r^2} \ .  
% \\
%\text{and:} \ \  \tfrac{1}{2r} \Psi(r)\cdot m_0 = \Phi(r)  \ ,  
%\text{and:} \ \  \tfrac{1}{2r^2} |\Psi(r)|^2 \int_0^{2\pi}g'(\theta)^2d\theta = \Phi^2(r)  \ ,  
%\label{eq:slaving-2}
\end{gather}
Here, $\srr$ is given by Eq.~(\ref{eq:TFT-1}) and $\sqq$ acts %may be interpreted 
as a Lagrange multiplier that enforces the 
condition (\ref{eq:slaving}), analogous to the inextensibility constraint underlying one-dimensional (1D) {\emph{elastica}}. 
%subject to the ``slaving'' constraint (\ref{eq:slaving}). 

%%%%%%%%%%%%%%%%%%%%%%%%%%%%%%
%%%%%%%%%%%%%%%%%%%%
%%%% FIGURE 3 %%%%%%%%%%%%%%%%
\begin{figure}
\centering 
\begin{center} 
\includegraphics[width=0.45\textwidth]{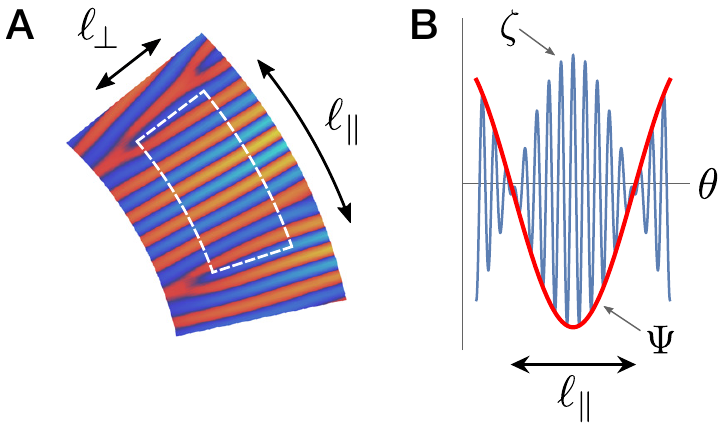} 
\end{center}
%\vspace{-0.1cm}
\caption{
(A) Schematic of a narrow, defect-free annular zone, where Eq.~(\ref{eq:ansatz-multi}) holds. 
%The width $\elld$ is the radial distance between defects in (same or adjacent) defect-rich buffers, and the azimuthal extent, $\elln$ (Eq.~\ref{eq:ell-amp-1}) is determined by energy-favored amplitude modulations.
(B) Schematic of the deflection $\zeta(r,\theta)$ at a given $r$, exhibiting rapid undulations of wavelength $\approx \barlambda$, modulated by the slowly-varying (complex) amplitude $\Psi(r,\theta)$.    
}     
%Morphological phase diagram, spanned by the parameters $\Bo$ and $\tau$ (and small values of $\epsilon$, such that $\epsilon/\Bo =Cst$), exhibits two qualitatively-distinct types of patterns in parameter regimes that are separated by a curve $\Bo_c(\tau)$. 
%, which may depend on the fixed ratio $\epsilon/\Bo$. 
%Above this curve we find in experiments and simulations patterns with a fixed number of radial wrinkles; below it we find defects-proliferated patterns with a fixed wavelength (well described by Eq.~\ref{eq:lambda-1}), that contains defect-rich zones separated by defect-free zones with strong amplitude modulations.}
\label{fig:3}
\end{figure}

Specializing to a narrow annulus, $r \in \rs \pm \elld$, where $\elld\ll \rs$ and $\Rin\!<\!\rs\!<\!L$ (Fig.~\ref{fig:3}), and expecting the effect of %the operator 
${\cal L}_1$ to vanish as $\Bo \to \infty$, 
%one may be motivated to  
%we may %first 
%ignore  it, 
%it seems natural to address this limit by ignoring 
%${\cal L}_1$ %in (\ref{eq:Fvk-1}), %Eq.~(\ref{eq:Fvk-1}), 
%such 
we notice 
%one may notice 
that Eqs.~(\ref{eq:slaving},\ref{eq:FvK-1})
describe a %are akin to 
collection of 1D, {\emph{decoupled}}  
%are equivalent to the 
{\emph{elastica} rings of length $2\pi r$, subjected to confinement $\Phi^2(r)$, Eq.~(\ref{eq:slaving}), and substrate 
%attached to foundation of 
stiffness $\Ksub$. 
%Using then One may thus use the 
%This observation motivates a ``monochromatic'', variable-separated ansatz:  
This motivates constructing a ``monochromatic'' ansatz from periodic {\emph{elastica}} solutions, labeled by an integer $m$:
%This observation motivates the construction of a ``monochromatic'' ansatz from the known solutions of the {\emph{elastica}} problem: 
 \begin{gather}
\zeta(r,\theta) \approx \Psi(r) g(\theta)  \nonumber \\ %; 
\ \text{with} \ g(\theta) = \cos(m\theta) \ ; \  \Psi(r) \!=\! {2r\Phi(r)}/{m} \ ,   \label{eq:ansatz-mono}
%\nonumber 
\\
{\rm where:} \ \ \sqq(r) \!= \!-[B(m/r)^{2} \!+\! K (m/r)^{-2}] \ . 
% \ \ , \ \ \text{with:} \ \ |g'(\theta)| \gg r |\Psi'(r)| \  , 
\label{eq:sqq-1}
\end{gather}
%and ``eigenvalue'' $\sqq(r) \!= \!-[B(m/r)^{2} \!+\! K (m/r)^{-2}]$. 
The resulting %associated 
energy of bending and substrate deformation is: 
\begin{gather}
\! U_m\! \approx %\tfrac{1}{2}
%\sqrt{B\Ksub} \!
\bar{C} + 
\frac{1}{2} B%m^2   
\int \!d\theta \! \int \! dr \ \Psi(r)^2 r^{-3} m^2 \left(m - \barm(r)\right)^2 
%\left\{1 + (m - \barm(r))^2) \right\}
\label{eq:energy-1}
%\end{gather} 
%\begin{equation}
\\
\text{\rm where : }  \ \ \ \ \ \barm(r) \equiv {2\pi r}/{\barlambda} \ \ ; \ \barlambda \equiv  2\pi \! \sqrt[4]{{B}/{\Ksub}} \  , 
%\ms \!\approx \! 2\pi \rs/\lambdas  \  ; \  
%{\lambdas} \!= 2\pi \! \sqrt[4]{\frac{B}{\Ksub}} \  ;   \ 
%\ms \!\approx \! \frac{2\pi \rs}{\lambdas}  \  ; \  
%\ms \!\approx \! 2\pi \rs/\lambdas  \  ; \  
%{\lambdas} \!= 2\pi \! \sqrt[4]{\frac{B}{\Ksub}} \  ;   \ 
\label{eq:m0}
\end{gather}
and $\bar{C}$ is a constant $\sim \sqrt{B\Ksub}$. % (see {\emph{SI}}). 
Minimizing $U_m$ 
over integers $m$ 
%in a %sufficiently 
%narrow annulus
%, $\elld \!\!\ll \!\!\rs $, 
yields
%Assuming a sufficiently narrow annulus, $\elln \!\!\ll \!\!\rs $, minimizing $U_m$ yields % in each narrow annulus yields 
%implies that for $r \in \rs \pm \ellt/2$ is minimized by 
$m \!\approx\! \ms = 2\pi \rs/\barlambda$, %thus 
recovering Eq.~(\ref{eq:lambda-1}) for $\Bo \!\to\! \infty$.    
%Minimizing $U_m$ one 
%recovers 
%Eq.~(\ref{eq:lambda-1}) for $\Bo \to 0$, as long as $\elln/\rs \ll 1$ in that limit. 

The argument thus far is merely a reformulation of  
%merely reformulates is merely a reformulation of 
previous analyses of the wrinkle wavelength \cite{Cerda03,Paulsen16}, underscoring  
%but it}
%Nevertheless, 
%{\color{blue} this argument -- which is merely a reformulation of previous analysis of the wrinkle wavelength \cite{Cerda03,Paulsen16} -- }  
%conceals 
two %profound %intimately related 
difficulties. First, 
%the integer $m_0$ must change between consequent narrow annulli,  however 
the ansatz (\ref{eq:ansatz-mono}) does not indicate how transitions occur 
%a mechanism for the necessary transition 
between distinct %integer 
values of $\ms$ at adjacent narrow annuli. Second, %attempting to develop the asatz (\ref{eq:ansatz-mono}) into an expansion in $Bo$, 
%attempting to re-introduce ${\cal L}_1$ and 
%expanding 
for finite $\Bo \!\gg\! 1$ %, one finds that 
the perturbation imposed by the operator ${\cal L}_1$ is {\emph{resonant}} ({\emph{i.e.}} ${\cal L}_1 \zeta \propto g(\theta) = \cos(\ms \theta)$ is a zero mode of ${\cal L}_0$), 
akin to periodically driving a harmonic oscillator at its resonant frequency. Hence, Eq.~(\ref{eq:FvK-1}) is impervious to regular expansion around the ansatz (\ref{eq:ansatz-mono}) unless ${\cal L}_1\zeta \!=\!0 \Rightarrow \Psi''(r) \!=\! 0$, which is incompatible with 
%clearly in conflict with 
Eq.~(\ref{eq:slaving}).   

Motivated by %our observations in Fig.~{\ref{fig:2}b,d, and by 
%a formal analogy to 
multi-scale perturbation theory of nonlinear dynamics problems\footnote{Such problems often feature %affect    
%Resonant perturbation is common in nonlinear dynamics problems, whereby the failure of a simple perturbation theory 
%signals the  
%it affects a 
%emergence of 
a slow temporal modulation of the amplitude of %an otherwise 
a periodic signal (\emph{e.g.}, $ A_0 \cos[\omega_0 t] \to \text{Re}[A(t) e^{i (\omega_0 +\Delta \omega) t} ]$, such that $|\Delta \omega| \ll |\frac{A'(t)}{A(t)}| \ll \omega_0$).}, we generalize the ansatz (\ref{eq:ansatz-mono}) to: 
%\begin{gather}\zeta(r,\theta) \approx \text{Re} [\Psi(r,\ttheta) e^{i\ms\theta}]  \  ,  \ \text{where:}  \ \ttheta \equiv  \ms \ \theta/\sqrt{\Bos} \label{eq:ansatz-multi} \\ \text{and:} \  \Bos \!=\! {\Ksub {\rs}^2\!/\!\srr(\rs)}  \!=\!\Bo (\rs \!/\! \Rin)^{3} \gg 1  \ ,\end{gather}
\begin{gather}
\zeta(r,\theta) \approx \text{Re} [\Psi(r,\theta) e^{i\ms\theta}]   \ , 
\label{eq:ansatz-multi}
\end{gather}
where the complex amplitude $\Psi(r,\theta)$ satisfies $|\partial_\theta \Psi|\! \sim\! 
{\Bo}^{-b} \ms |\Psi | \!\ll\! \ms |\Psi |$ for some $b\!>\!0$.
%exponent $b\!>\!0$.   
%%%%
%The basic idea here -- expanded upon in {\emph{SI}} --
%is obtaining a revised solvability condition that guarantees the avoidance of resonant terms in solving (\ref{eq:FvK-1}) for finite $\Bo \gg 1$, thereby making the generalized ansatz~(\ref{eq:ansatz-multi}) a feasible approximant for a certain class of amplitude functions. % $\Psi(r,\theta)$.       
%({\emph{i.e.}} avoidance of resonance) 
%
%In {\emph{SI}} we explain in detail how the generalized ansatz~(\ref{eq:ansatz-multi}) remedies the resonant perturbation ${\cal L}_1$, providing a mathematically-valid approximant of the wrinkle pattern for finite $\Bo \gg 1$. 
%Here, we only seek to elucidate how it underlies the transitional mechanism amongst integer values of $\ms$, and thereby the mesoscale structure exhibited in our experiments and simulations (Fig.~{\ref{fig:2}b,d).       
%For that purpose, it is useful 
%The basic idea is to assume 
%Assuming the generalized ansatz~(\ref{eq:ansatz-multi}) with $|\partial_\theta \Psi| \sim  {\Bo}^{-b} \ms |\Psi |$ (where $b>0$), 
%such that the solution of Eq.~6 in the limit $\Bo \to \infty$ is compatible with the ansatz (\ref{eq:ansatz-multi}) for any $b>0$, 
%as long a solvability condition ({\emph{i.e.}} avoidance of resonance) is satisfied:  
For the %generalized 
ansatz~(\ref{eq:ansatz-multi}), the avoidance of resonant effects in a perturbative expansion of Eq.~(\ref{eq:FvK-1}) around the limit $\Bo \!=\!\infty$ implies the equation: % an ``amplitude equation'': 
%solution to (for finite $\Bo \gg 1$) 
%solvability condition becomes: 
%and the solvability condition for an expansion around it for finite $\Bo \gg 1 $ becomes:    
%\begin{equation}
%$[\srr \tfrac{\partial^2}{\partial r^2} + {\Bo}^{-2b} \tfrac{4}{r^2}\tfrac{\partial^2}{\partial \theta^2}]\Psi(r,\theta) = 0 $.   
%Since $\Bo \gg 1$, 
%The only nontrivial %value of the 
%exponent $b$ %for which 
%that balances
%azimuthal and radial derivatives of the amplitude for $\Bo \gg 1$
%are in balance at $\Bo \gg 1$, 
%in this limit
%is $b\!=\!1/2$, yielding: 
\begin{equation}
[\srr \tfrac{\partial^2}{\partial r^2} + 4|\sqq| \tfrac{1}{r^2}\tfrac{\partial^2}{\partial \theta^2}]\Psi(r,\theta) = 0  \ ,
\label{eq:Laplace}
\end{equation}
such that the only nontrivial value of the 
exponent $b$, for which 
the two terms in Eq.~(\ref{eq:Laplace}) are in balance  
%that balances
%the azimuthal and radial derivatives of the amplitude 
for $\Bo \gg 1$, 
%are in balance at $\Bo \gg 1$, 
%in this limit
is $b\!=\!1/2$ %, yielding: 
%(In {\emph{SI} we elaborate on the irrelevance of $b\neq 1/2$). 
(where we used Eqs.~(\ref{eq:sqq-1},\ref{eq:m0})). 
In contrast to the %monochromatic 
ansatz (\ref{eq:ansatz-mono}), Eq.~(\ref{eq:Laplace}) admits azimuthally-oscillatory solutions to the wrinkle amplitude. %$\Psi \sim e^{-2 \pi r/\ellt} \cos(\pi r \theta/\ellt}$,   
%$\Psi \sim e^{\pi r (2 + i\theta)/\ellt} $, %\cos(\pi r \theta/\ellt}$,   
%of the generalized ansatz (\ref{eq:ansatz-multi}), 
%and thereby a %thus gives rise to
Crucially, this facilitates a mechanism for transitioning between distinct integer values of $\ms$ at adjacent annuli: defects can nucleate within %whereby 
amplitude-suppressed zones ($\Psi(r,\theta)\! \approx\! 0$) at negligible energy cost. 
%We propose that such oscillatory solutions of Eq.~(\ref{eq:Laplace}) underlie the mesoscale structure observed in Fig.~?b and d. 

%Focusing on 
Within a narrow, defect-free annulus around \mbox{$r\!=\!r_a$}, an azimuthally-oscillatory solution to Eq.~(\ref{eq:Laplace}) satisfies:  
%may be approximated by: 
\mbox{$\Psi \!\propto\! \exp{[-{2 \pi}(r-r_a) \ellt^{-1} \sqrt{{|\sqq|}/{\srr(r_a)}}]} \cos({\pi r_a} \theta/{\ellt})$}. %and 
%An %rough   
A crude 
%Assuming a small defect-free zone of size $\ellt \gg \barlambda$ and $\elln$ centered at a radial distance $r_a$    
estimate of %the scale 
$\elln$, 
over which the amplitude varies azimuthally, 
may be obtained by requiring $\partial\Psi/\partial r  \sim \Psi/r_a$ to ensure compatibility with Eq.~(\ref{eq:slaving}), and recalling Eqs.~(\ref{eq:sqq-1},\ref{eq:m0}):    
% balancing the two terms in Eq.~(\ref{eq:Laplace}): 
%of azimuthal oscillations of the amplitude: 
\begin{equation}
\elln \sim \ellBTs \cdot r_a/ \barlambda 
\ \ \ \ {\rm where:} \ \ \ \ellBTs  \equiv \sqrt{B/\srr(r_a)} \ ,  
\label{eq:ell-amp-0}   
%{\rm where:}  \ \ell_{BC}^*(\bfx) = \sqrt{{\sigmaaux(\bfx)}/{B}}  \ ,   \nonumber  \\
%\ \  {\ellbends}(\bfx)  = \left( |\nabla \cdot \nsigma(\bfx)| + |{\partial_{\perp}\Phiaux (\bfx)}/{\Phiaux(\bfx)}|\right)^{-1}  \ .  \nonumber 
%\kappa_g(\bfx)  = {\ell_{BC}^*(\bfx)}/{\lambda_0}  \  ;  \  
%{\rm and:} \  
\end{equation}
%where we assumed both terms in Eq.~(\ref{eq:Laplace}) are of comparable magnitude, 
where $\ellBTs$ is 
%we employed Eqs.~(\ref{eq:sqq-1},\ref{eq:m0}), estimated $\partial^2\Psi/\partial r^2 \sim \Psi/r_a^2$, and introduced 
a (local) ``bendo-capillary'' length %$\ellBTs$ 
\cite{Schroll13}. %$ \ellBTs  = \sqrt{B/\srr(r_a)}$.  
%     
%\lambda_0 \approx 2\pi (B/\Ksub)^{1/4} \nonumber 
%(local) ``bendo-capillary'' length \cite{Schroll13}, 

%\section*{Energetic considerations for a modulated amplitude}
\section*{Coarse-grained energy and amplitude modulations}
%Energetic considerations for a modulated amplitude}
% and a covariant formula for $\elln$}
%
% by assuming both terms in Eq.~(\ref{eq:Laplace}) are of comparable magnitude, and           
%
%In order to identify the physically-relevant states (among the many azimuthally-oscillatory solutions) of Eq.~(\ref{eq:Laplace}), we 
%Before turning to analyze our %compare this prediction with %analyzing 
%experiments and simulations, we 
Let us now elaborate on the energetic hierarchy imparted by 
the ansatz (\ref{eq:ansatz-multi}), and derive a generic, quantitative version of Eq.~(\ref{eq:ell-amp-0}).     
Exploiting the separation of scales between the wrinkly undulations and their slowly-varying amplitude, we derive in the {\emph{SI}}
a Ginzburg-Landau-like energy functional for the amplitude $\Psi (r,\theta)$, by averaging out the FvK energy over the small scale 
$\barlambda$. 
%Focusing on a defect-free zone centered around $r_a$, 
%(whose size is assumed to be much smaller than $r_a$ yet sufficiently larger that $\Bo^{-1/2} \barlambda$), 
This coarse-graining calculation recovers the energy $U_m$ (\ref{eq:energy-1}), and yields two other terms: % (see SI): 
\begin{equation}
\!\!\Unonlin \!\approx \frac{Y}{2}\!\!\int \!\!\!d\theta \!\! \int\!\! \! rdr \  [\tfrac{m^2|\Psi|^2}{4r^2} \!-\!\Phi^2(r)]^2  \!+\! \tfrac{m^2}{4r^4}|\Psi|^4 (\tfrac{\partial\arg\Psi}{\partial\theta})^2   \ , 
\label{eq:energy-nonlin}
\end{equation}
%
%\noindent
%\vspace{-0.2cm}
\begin{equation}
%%%%(\partial_\ttheta \Psi)^2 + (\partial_r\Psi)^2   \ , 
%\!\!U_{\Psi}\! \approx \!\frac{\sqrt{B\Ksub}}{2{\rs}^2} \! \int_{\rs- \ellt/2}^{r+\ellt/2}\!\!\!\!\!\!\!\! \!\!rdr \! \left(\!\Bos {\ms}^2 |\partial_r\Psi|^2 + \tfrac{4}{r^2}|\partial_\theta \Psi|^2 \!\right)  , 
\!\!\!\!\!U_{\Psi}\! \approx \tfrac{1}{2}  \int \!d\theta \! \int \! rdr \ \left(
%\! \int_{\rs- \ellt/2}^{r+\ellt/2}\!\!\!\!\!\!\!\! \!\!rdr \! 
\srr |\partial_r\Psi|^2 + 
4 |\sqq|\ \tfrac{1}{r^2}|\partial_\theta \Psi|^2 \right) \  , 
\label{eq:energy-2} 
%\\
\end{equation}
where $\Phi, \srr, \sqq$ are given by Eqs.~(\ref{eq:TFT-1},\ref{eq:slaving},\ref{eq:sqq-1}). %,\ref{eq:m0}). 
The sum
%Taken together, 
$\Unonlin + U_m+U_\psi$ describes the {\emph{deviation}} of the energy of an actual wrinkle pattern from the TFT limit value (\ref{eq:Udom}), reflecting a hierarchy of energetic costs. %The term 
$\Unonlin$ (\ref{eq:energy-nonlin}) includes quartic terms in $\Psi$, which must be considered since %any nonzero values there 
they
are penalized by a large stretching modulus, $Y$, reflecting strain in excess of the residual level already accounted for in the limit value (\ref{eq:Udom}). 
The term $U_m$ (given by Eq.~\ref{eq:energy-1} above), %that we already discussed above 
reflects a balance between %the energies of 
bending and substrate (\emph{i.e.}, liquid gravity) energies, yielding
%which give rise to 
the favored wrinkle wavelength, $\barlambda$. 
%and its minimization determines an optimal wrinkle number $m(r) \approax m_a$ in a narrow annulus centered at $r=r_a$. 
The last term, $U_\psi$ reflects the energy cost for amplitude modulations 
and thereby the emergence of defects to enable %$\lambda(r) \approx \barlambda$ 
proximity 
of the wavelength $\lambda(r)$ to $\barlambda$
throughout the pattern. %(\ref{eq:m0}). 
%through which the wavelength remain close to $\barlambda$ throughout the pattern. 
%The Euler-Lagrange equation derived from $U_\psi$ is exactly Eq.~(\ref{eq:Laplace}). 
Its typical value is $\mathcal{O}({\Bo}^{-1})$ relative to the energy incurred by $U_m$, and its corresponding Euler-Lagrange equation is precisely Eq.~(\ref{eq:Laplace}). 

For a small, defect-free zone of azimuthal and radial extents $\elln,\elld \ll r_a$, the  
%Assuming a small defect-free zone of azimuthal extent $\barlambda \ll \ellt \ll r_a$ and radial extent $\ellt \ll r_a$ centered at a radial distance $r_a$ (see Fig.?), in which 
%$m(r) \approx m_a$ (such that $U_m$ is as close as possible to its minimum), the optimal, azimuthally-oscillatory solution to Eq.~(\ref{eq:Laplace}) must be therefore obtained by considering the energy $\Unonlin$ (\ref{eq:energy-nonlin}). The 
second term in Eq.~(\ref{eq:energy-nonlin}) indicates that strain is induced by 
any deviation of wrinkles from the tension-carrying lines ({\emph{i.e.}}, radial lines, for which 
$\tfrac{\partial\arg\Psi}{\partial\theta} = 0$). %yields strain, 
Hence, the radial orientation of wrinkles persists in defect-free zones,  
%({\emph{i.e.}} $\tfrac{\partial\arg\Psi}{\partial\theta}\approx 0$), 
locally suppressing the smectic order imparted by uniformly-spaced wrinkles (for which $\tfrac{\partial\arg\Psi}{\partial\theta}\approx \barm(r) - m_a$). 
%Consequently, we focus our analysis on real-amplitude azimuthally-oscillatory solutions of Eq.~(\ref{eq:Laplace}): $\Psi(r,\ttheta) = c \ e^{\pm 2b(r-\rs)} \cos(b r \ttheta)$, where the magnitude $c$ and period $b = \pi/\elln$ are found by considering the first term in (\ref{eq:energy-nonlin}). 
In contrast, % to the second term, 
the first term in Eq.~(\ref{eq:energy-nonlin}) does not vanish for any azimuthally-oscillating amplitude, but its cost can be made negligible
(%more precisely, 
energy density $\sim Y(\elld/r_a)^4$),   
%. Nevertheless, as long as $\ellt/r_a \ll 1$ ({\emph{i.e.}} high density of defects that is in turn enabled by their own low energetic cost in amplitude-suppressed zones), the resulting strain can be made negligible 
by requiring the corresponding integrand 
%square-bracketed integrand in Eq.~(\ref{eq:energy-nonlin}) 
and its radial derivative to vanish upon integrating over an oscillatory period of $\Psi$, yielding: %, providing two conditions that determine the constants $b,c$. 
%This analysis yields:
% a quantitative version Eq.~(\ref{eq:ell-amp-0}):  
\begin{gather}
%\elln \approx k_g \cdot \ell_{\rm bend} \label{eq:ell-amp}  \\ 
\!\!\elln \approx  4\pi^2 \ellBTs \ellbends/\barlambda   \ \ \ {\rm where} 
%\frac{\ellbends}{\barlambda}  \ \ \ {\rm where} 
%{\rm where:}  \ \ell_{BC}^*(\bfx) = \sqrt{{\sigmaaux(\bfx)}/{B}}  \ ,   \nonumber  \\
%\ \  {\ellbends} \!=\! \left( \frac{1}{r_a} + |\frac{\Phi'(r_a)}{\Phi(r_a)}| \right)^{-1}  \!\! .  
\ \  {\ellbends}^{-1} \!=\! \frac{1}{r_a} + |\frac{\Phi'(r_a)}{\Phi(r_a)}|   . %\nonumber 
\label{eq:ell-amp-1}  
%|{\partial_{\perp}\Phiaux (\bfx)}/{\Phiaux(\bfx)}|\right)^{-1}  \ .  \nonumber 
%(\bfx)  = \left( |\nabla \cdot \nsigma(\bfx)| + |{\partial_{\perp}\Phiaux (\bfx)}/{\Phiaux(\bfx)}|\right)^{-1}  \ .  \nonumber 
%\kappa_g(\bfx)  = {\ell_{BC}^*(\bfx)}/{\lambda_0}  \  ;  \  
%{\rm and:} \  
\end{gather}
In addition to replacing the scaling relation in Eq.~(\ref{eq:ell-amp-0}) by a number 
%with a numerical pre-factor 
($4\pi^2$), the length $r_a$ %in Eq.~(\ref{eq:ell-amp-0}) 
is replaced by yet another        
%\lambda_0 \approx 2\pi (B/\Ksub)^{1/4} \nonumber 
local length $\ellbends$, which derives from the (planar) curvature of the axis $\hat{\theta}$ along which wrinkles 
% the wrinkly undulations 
 suppress an imposed compression.

\section*{General form of theoretical results} 
%Beyond the Lam\'e set-up, the 
The arguments underlying Eq.~(\ref{eq:ell-amp-1}) apply to  
%
%The dependence of $\ellt$ on the wrinkle wavelength $\barlambda$ and the lengths $\ellBTs ,  \ellbends$, both of which are determined by  
%the TFT solution of the Lam\'e problem (Eqs.()), suggests that Eq.~(\ref{eq:ell-amp-1}) to 
a broad class of confinement problems, for which a  
wavelength $\barlambda(\bfx)$ is favored in the vicinity of a point $\bfx$ on the sheet by competition of  
%the competition of 
bending resistance and substrate-induced stiffness (akin to Eqs.~\ref{eq:lambda-1},\ref{eq:m0}). 
%Indeed, a more general formulation of Eq.~(\ref{eq:ell-amp-1}) can be obtained by assuming 
For given boundary loads and substrate shape, a TFT solution may be found analytically or numerically 
\cite{Roux18}, yielding 
%Assuming that a TFT solution %of such a confinement problem 
%is known and yields 
the macro-scale fields -- a %bent 
director $\naux(\bfx)$ along which wrinkles  
``waste'' a fraction of arclength
%(= \theta$ in the Lam\'e problem) 
%excess-length-per-wrinkle 
$\Phiaux^2(\bfx)$, and a tensile stress $\sigmaaux(\bfx)$ (akin to Eqs.~(\ref{eq:slaving},\ref{eq:TFT-1}), respectively). 
Here, ``aux'' refers to an auxiliary state that describes %characterizes the problem in 
the {\emph{singular}}, infinite bendability limit, % ($\epsilon=0$)
%realized only when the confined body has zero bending modulus, 
of a hypothetic body with zero bending modulus,  
and ``$\parallel,\perp$'' denote the (curvilinear) planar axes, along and normal to $\naux(\bfx)$. %For such a confinement problem, 
Our analysis predicts that if $\naux(\bfx)$ is bent ($\nabla \times \naux \neq 0$) then  
a defect-riddled pattern consists of defect-free domains where $\lambda(\bfx) \approx \barlambda(\bfx)$,    
%for which the wavelength $\lambda(\bfx) \approx \barlambda(\bfx)$ almost everywhere, consists of defect-free domains 
whose longitudinal scale, $\elln$, is given by Eq.~(\ref{eq:ell-amp-1}), with:  
\begin{gather}
%\elln \approx k_g \cdot \ell_{\rm bend} \label{eq:ell-amp}  \\ 
%\elln \approx  4\pi^2 \ellBTs \cdot \ellbends/\barlambda \label{eq:ell-amp-2}  \\ 
%{\rm where:}  \ 
\ellBTs  \!=\! \sqrt{\frac{B}{\sigmaaux(\!\bfx\!)\!}}  \ ,   %\nonumber  \\
%  {\ellbends}  \!=\! 
%\frac{1}{( |\nabla \!\times\! \nsigma(\!\bfx\!)| \!+\! |\frac{\partial_{\perp}\Phiaux (\!\bfx\!)}{\Phiaux(\!\bfx\!)}|)}  \ . %^{-1}   . 
  {\ellbends}^{-1}  \!=\! 
{ |\nabla \!\times\! \nsigma(\!\bfx\!)| \!+\! |\frac{\partial_{\perp}\Phiaux (\!\bfx\!)}{\Phiaux(\!\bfx\!)}|}  . %^{-1}   . 
 \label{eq:ell-amp-2}
%\nonumber 
%\kappa_g(\bfx)  = {\ell_{BC}^*(\bfx)}/{\lambda_0}  \  ;  \  
%{\rm and:} \  
\end{gather}

The applicability of our result to another axial confinement problem is demonstrated
%applicability of Eqs.~(\ref{eq:ell-amp-1},\ref{eq:ell-amp-2}) for another (albeit still axial) confinement problem is shown  
in Fig.~\ref{fig:1}c, which shows %in grayscale 
wrinkles %the wrinkles %pattern 
near  
%the azimuthal profile of the wrinkle amplitude at 
the edge of a circular sheet of radius $W$ attached to a spherical Winkler foundation (a ball of springs of stiffness $\propto \Ksub$ 
and rest length $R\gg W$ \cite{Hohlfeld15,Davidovitch19}). For that problem, the tensional bendability $\epsilon^{-1}$ and Bond number $\Bo$ are defined similarly to Eq.~(\ref{eq:nondim-param}), with $\gin$ replaced by 
the tensile load $\gamma$ exerted at the %sheet's 
perimeter, and the parameter analogous to $\tau$ 
%whose increasing 
%in that problem
%that rise to azimuthal confinement %(akin to $\tau$ in Eq.?) 
is $\alpha = \tfrac{YW^2}{\gamma R^2}$, which controls the strength of azimuthal confinement. 
A TFT solution (Sec.~IV.A-C in Ref.~\cite{Hohlfeld15}) yields   
%of that problem yields 
expressions for $\srr(r)$ and $\Phi(r)$ 
%(Sec.~IV.A-C in Ref.~\cite{Hohlfeld15}) 
analogous to Eqs.~(\ref{eq:TFT-1},\ref{eq:slaving}), which we substitute in Eq.~(\ref{eq:ell-amp-2}) to compute $\elln$ at $r=W$ (scale bar in Fig.~\ref{fig:1}c).
%the  shows the value of $\elln$ at the perimeter ($r=W$) 
%predicted by Eqs.~(\ref{eq:ell-amp-1},\ref{eq:ell-amp-2}). %(Eq.~\ref{eq:ell-amp-1}) 
%the predicted value 
%(A detailed study of this system is deferred to a separate publication). 
}

\section*{Comparison with experiments and simulations} 

%In Fig.~\ref{fig:4}, we compare our predictions with experiments and simulations. %Figure~\ref{fig:4} shows the maximal amplitude at each latitude (radius $r$), as a function of $r$, extracted from simulations for a few values of the confinement parameter $\tau$~(Eq.~\ref{eq:nondim-param}).
Figures~\ref{fig:4}a,\ref{fig:4}c show the azimuthal profile of a sheet with $\Bo > \Bo_c$ at a single radius, which we quantify using the image intensity $I(\theta)$ in experiments~\cite{Paulsen16}, and the vertical deflection $\zeta(\theta)$ in simulations.
A pronounced %dominant 
wrinkle wavelength is clearly present, with large amplitude modulations. 
Figures~\ref{fig:4}b,\ref{fig:4}d show the measured lengthscale of these modulations at several radii (%in units of 
normalized by {the 
measured %observed 
$\lambda$}).  
%We measure the lengthscale of these modulations at several radii (in units of {the observed $\lambda$}), as shown in Figs.~\ref{fig:4}b,\ref{fig:4}d. 
%The values and the radial dependence are described well by our theory with no free parameters [Eq.~(\ref{eq:ell-amp-1})]. 
Figure~\ref{fig:4}e compares %the 
measured and predicted values of $\ell_{||}/\lambda$ %$\ell_{||}/\Rin$ 
for several %sheets with varied  
values of $\Bo$ and $\tau$.  %and $\epsilon$, 
No fitting parameters are used. 
The experiments exhibit %are in 
good agreement with the theory, whereas the simulation results are systematically lower than expected and approach 
the predicted values only at large $\Bo$ and small $\tau$, where the extent of the wrinkled zone is relatively small (Eq.~\ref{eq:TFT-1}). 
We attribute this discrepancy to the numerical difficulty in reaching the most low-lying energy states. %due to the multiple hierarchies of energies in the problem. 
In particular, at large $\tau$ we find many metastable states where defects are scattered throughout the wrinkled region, thereby decreasing the characteristic distance $\elln$ between the defect-rich zones.
%We interpret these as metastable states with higher system energy. 
%When $\tau$ is smaller, we are able to attain states where the defects gather into a smaller number of defect-rich zones; this lowers the system energy and increases the characteristic distance $\elln$ between the zones. %(as shown by the different symbols in Fig.~\ref{fig:4}e.

%
%
%\jp{The simulation results are systematically lower than the theory, but they approach the prediction at large $\Bo$ and small $\tau$ where we have more robust techniques for minimizing the elastic energy. 
%Although the measurements in the simulations are systematically lower than the prediction, 
%Finding low-lying energy states is complicated by the separation in energy scales between the terms dictated by stretching, bending, and amplitude modulations. (These disparate energy scales set the wrinkle extent $L$, the wrinkle wavelength $\lambda$, and the modulation lengthscale, $\elln$, respectively.) We thus often observe metastable states where the wrinkle defects are scattered throughout the film, rather than gathering in a smaller number of defect-rich zones, %separated by smectic regions of parallel wrinkles, 
%which increases $\elln$ towards the predicted value. } 
%(with similar values of the control parameters $\tau, \epsilon, \Bo$), and their Fourier spectrum (Figs.~\ref{fig:4}b,\ref{fig:4}d, respectively). 
%
  
%analysis of the mesoscale structure in this system is deffered to a separate publication).  

%Turning now to 
Considering the phase diagram (Fig.~\ref{fig:2}i), we note that a transition between defect-proliferated and defect-free patterns can be rationalized by comparing the energies $U_m$ (\ref{eq:energy-1}) and $U_{\Psi}$ (\ref{eq:energy-2}),  
%of these two morphological types, 
whose respective scalings with the length of the wrinkled zone ($L-\Rin \!\sim\! \tau-2$) can be 
found for $0<\tau-2\ll 1$ (see {\emph{SI}}). The energy $U_m$ of a defect-free pattern (described by Eq.~(\ref{eq:ansatz-mono}) such that $m(r) = m_0 \neq \barm(r)$) is larger than that of a defect-proliferated pattern (where $m(r) \approx \barm(r)$). The relative energy gain for defect proliferation is $\Delta U_m \sim \Ksub\Rin^2 (\tau-2)^4$. In contrast, the azimuthal modulations of the amplitude in a defect-proliferated pattern, which are absent from a defect-free pattern, %(described by Eq.~(\ref{eq:pure-states}), 
%entail a relative energy cost, $\Delta U_\psi \sim \gin (\tau-2)$. Although these estimates are specialized for $0<\tau-2\ll 1$, whereas the data in Fig.~\ref{fig:2}i is for $(\tau-2)>1$, the scaling relation obtained by comparing  $\Delta U_m$ and $\Delta U_\psi$ suggests a transition curve $\Bo_c(\tau) \sim (\tau-2)^{-3}$, in accord with our observation (Eq.~\ref{eq:Boc}).          
entail a constant relative energy cost $\Delta U_\psi \sim \gin$ as $\tau \to 2$. These estimates thus predict the scaling $\Bo_c(\tau) \sim (\tau-2)^{-4}$. 
%The difference from Eq.~(\ref{eq:Boc}), which is obtained by analyzing data in the regime $(\tau-2)>1$ (Fig.~\ref{fig:2}i), may be because these estimates assume $0<\tau-2\ll 1$. 
Although this exponent differs from the measured value of $-3.4 \pm 0.3$ in Eq. (3), our theoretical estimates assume $0 < \tau - 2 \ll 1$, whereas the data are for $(\tau - 2) > 1$.
%attributed to the fact that 
%A possible reason for the difference from the observed scaling (Eq.~\ref{eq:Boc}) is that the 
%data in Fig.~\ref{fig:2}i is obtained in the parameter regime $(\tau-2)>1$, whereas our 

%amplitude modulations (since enable the formation of defects at amplitude-suppressed zones), the energy of the of the amplitude-modulated anstaz is governed by the deviation of        
  
%Considering The width of the wrinkled annulus is $\Rin (\tau/2 -1)$, which we assume to be much larger than the characteristic radial distance $\ellt$ between defects,  $\gg \ellt$, such that a defect-proliferated stated consists of many narrow annuli separated by defects that enable the wrinkle    

%consider then the wrinkled annulus, of width $\approx \tau -2$, 
%Considering first (\ref{eq:ansatz-mono},\ref{eq:m0}), we note that the energy $U_m$ () is not optimal, since the wrinkle number is fixed ($m(r) = m_0$), and thus deviates from minimal 

%%%%%%%%%%%%%%%%%%%%
%%%% FIGURE 4 %%%%%%%%%%%%%%%%
\begin{figure*}
\centering 
\begin{center} 
\includegraphics[width=\textwidth]{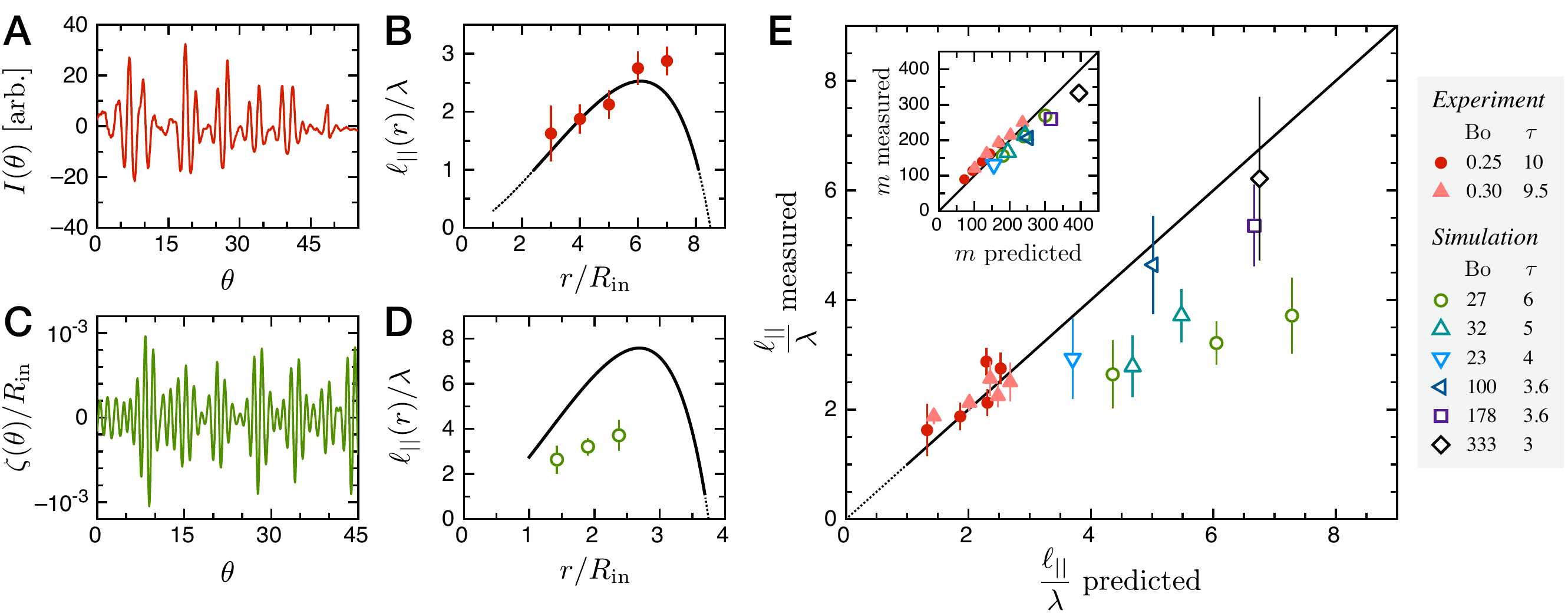} 
\end{center}
\caption{
Amplitude modulations and their lateral lengthscale, $\elln$ (normalized by the wavelength $\lambda$).
%/\lambda$, in experiment and simulation. 
(A) Image intensity versus $\theta$ along a circle of radius $r=26$ mm from the experiment in Fig.~\ref{fig:2}B with $\Bo=0.25$ and $\tau=10$. 
The signal oscillates with a wavelength $\lambda$, and the amplitude oscillates over a longer scale, $\elln$. 
(B) Radial dependence of $\elln/\lambda$ measured in the same experiment (circles).  
%and the prediction of  
%which is in good agreement with theory, 
%Eq.~(\ref{eq:ell-amp-1}) (solid curve). 
(C) Vertical displacement $\zeta(\theta)$ at $r/\Rin = 2.4$ in a simulation with $\Bo=27$ and $\tau=6$. 
(D) Radial dependence of $\elln/\lambda$ extracted from the same simulation (circles). 
In (B) and (D) the solid curves show the prediction of  
%which is in good agreement with theory, 
Eq.~(\ref{eq:ell-amp-1}). % is shown solid curve).
%follows a similar trend to the theory. 
(E) 
\textit{Main:} Measured $\elln/\lambda$ versus the predicted value, given by Eq.~(\ref{eq:ell-amp-1}). 
For sheets subjected to larger $\tau$ (and hence having longer wrinkles), we measure $\elln/\lambda$ at multiple radii.
The experiments are in good agreement with the theory, whereas simulations agree 
%with no free parameters.  
%The data fall reasonably close to the dashed line, 
%The simulations agree 
at large $\Bo$ 
and small $\tau$ (see text). %where the energy minimization is more robust
Note that although the experimental system has $\Bo <1$, the local parameter, $\Bos=\Ksub \rs^2/\srr(\rs)$ %(\ref{eq:Boc}), 
whose inverse is the expansion parameter in our theory ({\emph{SI Appendix}}), is substantially larger than $1$ in most of the wrinkled zone.   
\textit{Inset:} The measured wrinkle number agrees with the gravity-dominated value, $\bar{m}(r) = r(\rho g/B)^{1/4}$ (solid line). %The points are measured 
Data in the inset are sampled at the same radii as in the main panel.
}     
\label{fig:4}
\end{figure*}

\section*{Geometric conflicts and defect-proliferated states} 
%Since the 1950's, superconductors and liquid crystals have provided an arena  for exploring inhomogenous, defect-proliferated phases in thermodynamic equilibrium. 
%A prototypical liquid crystal former that puts our study in a broader context is a thermodynamic ensemble of elongated molecules (``nematogen''). 
To place our study in a broader context, we consider a thermodynamic ensemble of elongated molecules (``nematogen''). % that form a liquid crystal. 
Upon cooling or increasing density, this prototypical system transforms from an isotropic liquid to a ``nematic'' phase, where the molecular axes are parallel on average. 
The corresponding order parameter, reflecting broken rotational symmetry, is a director field with a uniform ground state ($\hat{n}(\bfx) = \hat{n}_0$). 
Cooling further leads to a ``smectic-A'' phase, where molecules form %are arranged into 
uniformly spaced %parallel 
layers 
%separated by a distance %thickness 
%$\lambda_0$, 
with normals parallel to the molecular director. 
The broken translational symmetry %along $\hat{n}_0$ 
underlies a %(scalar) 
complex order parameter with ground state $\Psi(\bfx) = A_0 \exp [i \hat{n}_0 \cdot \bfx /\lambda_0]$, 
characterizing the average spacing ($\lambda_0$) and favorable magnitude ($A_0$)  
%where $\lambda_0$ ad $A_0$ are, respectively, 
%the average layer spacing and intensity  
%is the layer spacing  
%$A_0$ characterizes a favorable level 
of smectic order. 
%The elasticity of the smectic phase is described by the coarse-grained, Landau-de Gennes energy, which penalizes $|\Psi(\bfx)| \! \neq  \! A_0$, as well as $|\nabla \hat{n}({\bfx}) | \!  \neq \!  0$ and $\partial^2_{\parallel} \arg \Psi  \! \neq \!  0$ \cite{DeGennesBook}
The %coarse-grained 
Landau-de Gennes energy describes the elastic response of the smectic phase, penalizing $|\Psi(\bfx)|\!\neq\! A_0$, as well as $|\nabla \hat{n}({\bfx}) | \!\neq\! 0$ and $\partial^2_{\parallel} \arg \Psi \!\neq\! 0$ \cite{DeGennesBook}.

%However, 
Such a simultaneous satisfaction of favorable orientational and translational orders is impossible if the nematogen is chiral.
%, since 
Here, the nematic is replaced by a ``cholesteric'' phase, of a twisted director  
%where the director is twisted 
%whose ground state is described by a twisted director 
($\hat{n}(\bfx) \cdot [\nabla \times \hat{n}(\bfx)] = q_0$ on average, where $q_0$ derives from
the nematogen structure).
%= b_0 \hat{k}_0$, where the rotor $b_0$ characterizes the %level of 
%chirality imparted by the molecular structure). 
In contrast to the non-chiral case, 
%it is impossible to form uniformly spaced, macroscopically flat layers, along a twisted director.
it is impossible to stack flat, uniformly spaced layers  
%macroscopically flat 
%layers, 
along a twisted director.
%a stacking of uniformly spaced, macroscopically flat 
%layers along a twisted director is impossible.    
Invoking an analogy to the Abrikosov lattice of magnetic flux lines in a type-II superconductor, this geometric conflict was predicted to give rise to a %defect-proliferated,   
 {\emph{twist grain boundary}} (TGB) phase \cite{Renn88} -- an inhomogeneous ground state %of the Landau-de Gennes energy, 
 where planes populated with screw-type dislocations %in the layered structure 
 separate mutually-tilted smectic domains. 
By suppressing %the level of 
smectic order near %in the vicinity of 
localized grain boundaries ({\emph{i.e.}} $|\Psi(\bfx)| < A_0$), the    
TGB phase enables the molecular director to attain there its desired twist ($ \hat{n}(\bfx) \cdot [\nabla \times \hat{n}(\bfx)] \approx q_0$). %, yielding 
The result is a spatial structure of alternating smectically-satisfied domains separated by cholesterically-satisfied %(but smectically-defected) 
buffer zones.

The crucial role of the director in 3d liquid-crystal phases and 2d wrinkled sheets %-- enforced by the nematogen structure in 3d liquid-crystal phases and by the confining forces exerted on a 2d wrinkled sheet) 
motivates us to further develop the %a recent 
analogy between these systems \cite{Aharoni17}. 
%draw parallels between these two physical systems. 
%In the simple case, 
When the energetics favor a constant director, a defect-free layered structure may emerge:
%, namely, 
the smectic-A phase for non-chiral nematogen, or a parallel array of uniform wrinkles for a uniaxially-confined sheet. 
In the geometrically-conflicted case of a non-constant director 
%where a non-constant director is energetically favored 
(twisted for chiral nematogen and bent for an azimuthally-confined sheet), one finds an inhomogeneous, defect-proliferated structure that retains parallel layers in separated 
%defect-free 
domains: % , \emph{i.e.}, 
the TGB phase or a defect-rich wrinkled film. 

This analogy is bolstered by contrasting our coarse-grained energy,  
%the coarse-grained energy of our system, 
$U_m + U_{\Psi}  + \Unonlin$ (Eqs.~\ref{eq:energy-1},\ref{eq:energy-2},\ref{eq:energy-nonlin}), and that of 
chiral nematogen (Eq.~2.11 of Ref.~\cite{Renn88}). 
In our 2d wrinkled sheet, the energy $U_m$ favors smectic order, 
%of uniformly-spaced wrinkles 
whereas the confining forces favor a bent director, $\naux =\hat{\theta}$, 
through the energy $\Unonlin$. If the energetic penalty of $U_m$ is %too 
small, the director $\naux$ is imposed forcefully, precluding
%prevents the formation of 
smectic order (\emph{i.e.} $\lambda(r) \propto r$) in analogy with the cholesteric phase 
($\Bo < \Bo_c$ in Fig.~\ref{fig:2}i). In contrast, if $\Bo > \Bo_c$, %(\ref{eq:Boc}), 
proliferation of defect-rich, amplitude-suppressed zones enables a partial recovery of smectic order (\emph{i.e.}, $|\lambda(r) - \barlambda| \ll \barlambda$) in defect-free domains, in analogy with the TGB phase. % Hence, 
The parameter $\Bo$ is thus akin %similar 
%to the thermodynamic field (%proportional to 
%$\sim$ 
to a Frank modulus, which penalizes
deviations of the director's twist from the value $q_0$,  
%of the molecular director from the value favored  
%to %plays a similar role to 
%the cholesteric rotor 
%rotor magnitude 
%magnitude of a twisted director 
%$b_0$, 
imparted by a nemtogen %molecular 
chirality. % in 3D liquid crystals. 
Furthermore, the size (``coherence length'') of smectic domains in the TGB phase 
derives from the minimal energy
associated with varying the order parameter ($|\Psi|:0\to A_0$) between  
defect-rich planes and %its favorable level in 
defect-free domains.  
%the coherence length underlying %that underlies 
%the size of smectic domains in the TGB phase derives from the minimal energetic cost 
%for varying the magnitude $|\Psi|$ of the smectic order parameter  
%complex amplitude (the smectic order parameter) 
%from zero at defect-rich planes to the %an %finite, 
%energetically-
%favorable level $A_0$ 
%in defect-free domains. 
This is again similar to our system, where the azimuthal extent $\elln$ of defect-free zones (Eq.~\ref{eq:ell-amp-1}) derives from the %minimal 
energy ($U_\psi + \Unonlin$) required to generate regions where the wrinkle amplitude $|\Psi|$ is suppressed.}     
\section*{Discussion}
%Confinement of thin solids often generates %in them 
%compressive stresses whose optimal suppression is often obtained through 
%wrinkles -- highly curved ``arclength wasting'' undulations, whose wavelength vanishes with the body thickness and whose director is dictated by the confining forces.  
Wrinkle patterns -- highly curved periodic undulations that ``waste'' an excess length -- are common in strongly confined thin solids that are  
%For 
%strongly confined bodies, 
%whose gross shape %of the confined body
forced to reside close to %in the vicinity of 
a smooth substrate. For such problems, tension field theory or its recent extension \cite{Davidovitch19}  
%(or its extension to purely-geometric  
%confinement \cite{Davidovitch19}) 
predict a 
%a hierarchical energetic structure  
%pronounced hierarchical structure of the elastic energy 
%that underlies a 
macro-scale thickness-independent director field $\naux(\bfx)$, a micro-scale %thickness-dependent 
wavelength $\barlambda(\bfx)$ %that  
%emanates from a 
%balances %between 
%bending rigidity and substrate stiffness, %and vanishes with the body thickness, 
and a corresponding stress field.  
%of the confined body and the substrate stiffness. 
%Deviations of the actual director or wavelength from these values are penalized, respectively, through the stretching modulus ($Y$), or the geometric mean $\sqrt{B\Ksub}$ of the two other modulii. 
The simplest pattern -- a smectic-like array of uniformly-spaced, parallel wrinkles -- %exhibiting smectic-like order -- % due to uniaxial compression of a uniform sheet on a uniform substrate. 
%This response %in which case 
emerges when both $\naux(\bfx)$ and $\barlambda(\bfx)$ are constants. However, if either %of these %two
%confinement-imposed 
field is spatially-varying, it may be impossible for the confined body to satisfy both fields everywhere, and the ensuing negotiation gives rise to a host of mesoscale morphologies. 

Our work addresses %strong 
confinements characterized by a uniform 
$\barlambda$ and a purely bent director, 
%$\naux = \hat{\theta}$ (for which 
({\emph{i.e.}} $\nabla \!\cdot \naux \! =\!0, \nabla\! \times \!\naux \!\neq\! 0$). %Focusing on a prototypical example for such confinements, 
Studying the Lam\'e set-up as a prototypical model for such problems, 
we showed that the pattern may either consist of a fixed number of wrinkles, absent of smectic order, or be characterized by amplitude modulations over a mesoscale $\elln$ (Eq.~\ref{eq:ell-amp-1}) that enable proliferation of defect-rich zones and thereby partial recovery of smectic order (lower and upper parts of Fig.~\ref{fig:2}i, respectively). 
A different type of
%nother notable example of 
confinement with uniform $\barlambda$ and non-uniform but unbent director \mbox{($\nabla\! \times \!\naux \!\approx\! 0$)} yields a qualitatively different mesoscale structure, an example of which was realized by forcing %occurs when 
a %polygonal 
patch of a spherical shell 
%is forced 
to reside close to a plane \cite{Aharoni17,Tobasco19} (Fig.~\ref{fig:1}b). Here, the unbent director %$\hat{n}(\bfx)$ 
is piece-wise constant, containing %being %unbent ($\nabla\! \times \!\naux \approx 0$) but 
splay ($\nabla\! \cdot \!\naux \neq 0$) only at ``domain walls'' that separate defect-free, smectically-ordered domains of uniformly-spaced wrinkles \cite{Aharoni17}.
%, and whose %Here, it is those domain-walls, rather than  
%width %of such domain walls 
%may be obtained by analogy to the elasticity of a 2D smectic phase \cite{Aharoni17}. 
Two other notable 
confinement types may occur even under uniaxial compression ({\emph{e.g.}} $\naux = \hat{y}$), if the locally-favorable wavelength is spatially-varying such that $\nabla \barlambda \times \naux \neq 0$, %\cite{Brun19}, 
or the wrinkle amplitude itself is forced to vary spatially 
%due to boundary conditions 
such that $\nabla \Psi \times \naux \neq 0$. A recent experimental study that addressed the former case %of these confinement types 
employed a sheet with a non-uniform thickness (where patterns of defects that resemble Fig.~\ref{fig:2}c,d were observed) \cite{Brun19}, whereas the latter type, which underlies ``wrinkle cascades'' (Fig.~\ref{fig:1}a), was realized through amplitude-suppressing boundary conditions \cite{Huang10,Vandeparre11}.

While our theoretical analysis pertains to confinement problems with bent director fields ($\nabla\! \times \!\naux \neq 0$), we anticipate that the coarse-graining approach initiated in Ref.~\cite{Aharoni17} and further developed here may provide a unified framework to analyze mesoscale structures in a broad class of confinement problems \cite{Paulsen19}. At its core there is an energy functional 
%(\ref{eq:Um}.. 
of a slowly-varying complex %amplitude 
function, whose magnitude describes the wrinkle amplitude and whose phase describes deviations from an asymptotic (thickness-independent) director field $\naux$, imposed by the confining forces. Such a Ginzburg-Landau-like functional may be obtained by expanding the full elastic energy around a suitable TFT limit and coarse-graining over the wrinkle micro-scale. Pursuing this approach further      
%We believe that the coarse-graining approach will 
may reveal new analogies with %the theories of 
liquid crystals and superconductors, %beyond the examples addressed in \cite{Aharoni17} and our work, 
and elucidate the remarkable %morphological
complexity of wrinkle patterns. 
\section*{Materials and Methods}

\noindent\textbf{Experiments.}
%Experiments were performed using ultrathin polymer films that were manufactured on glass substrates and floated on a water bath. 
Polymer films were made by spin-coating solutions of polystyrene ($M_\text{n} = 99$k, $M_\text{w} = 105.5$k, Polymer Source) in toluene ($99.9\%$, Fisher Scientific) onto glass substrates %, as in
following Ref.~\cite{Huang07}. 
A white-light interferometer (Filmetrics F3) was used to measure film thickness, which was uniform over each film to within $4\%$. 
Methods for determining $\gin,\gout$, and descriptions of our image analysis routines are provided in the %{\emph{Supplementary Information (SI)}}. 
\emph{SI Appendix}.

%\subsection*{Film Preparation} 
%Films were made by spin-coating solutions of polystyrene ($M_\text{n} = 99$k, $M_\text{w} = 105.5$k, Polymer Source) in toluene ($99.9\%$, Fisher Scientific) onto glass substrates %, as in
%following Ref.~\cite{Huang07}. 
%A white-light interferometer (Filmetrics F3) was used to measure film thickness, which was uniform over each film to within $4\%$. 

%\subsection*{Determination of $\gout$ and $\gin$}
%The liquid-vapor surface tension, $\gamma_{lv}$, was measured with a Wilhelmy plate under typical experimental conditions, yielding $\gout=\gamma_{lv}$. 
%For each experimental image, the confinement ratio $\tau$ was deduced from the wrinkle length via a relation that was derived and validated previously for a finite annulus in the far-from threshold regime \cite{Pineirua11,Taylor15}. 
%Then, $\gin = \tau \gout$. 

%\subsection*{Wrinkle Analysis} 
%Measurements in Figs.~\ref{fig:2}e-h and \ref{fig:4}e of the wrinkle number, $m(r)$, were obtained using a custom automated image analysis following Refs.~\cite{King12,Paulsen16}. 
%After an initial filtering step to reduce noise and lighting gradients, an autocorrelation of the intensity versus $\theta$ was performed at each radius within a region free of material imperfections, effectively averaging over many wrinkles. The wrinkle number was extracted from this oscillating signal.

\medskip
\noindent\textbf{Simulations.}
Finite-element simulations were performed in ABAQUS, as detailed in the %{\emph{Supplementary Information (SI)}}. 
\emph{SI Appendix}.

%\bigskip
%\noindent\textbf{Data availability}
%The data that support the findings of this study are available in the {\emph{SI Dataset}}.
%from the corresponding authors upon reasonable request.}

\acknowledgements{
%We are grateful to 
We thank O. Agam, D. Bartollo, G. Grason, N. Menon, and R. Sknepnek for useful discussions; D. O'Kiely and D. Vella for a critical reading of the manuscript.    
Simulations were performed at the Triton Shared Computing Cluster at the San Diego Supercomputer Center and the Comet cluster (Award no. TG-MSS170004 to T.Z.) in XSEDE. Funding support from NSF-DMR-CAREER-1654102 (M.M.R. and J.D.P.), NSF-CMMI-CAREER-1847149 (J.C. and T.Z.), NSF-DMR-CAREER-1151780, NSF-DMR-1822439  
%and the W.M. Keck Foundation 
(O.T. and B.D.), is gratefully acknowledged. 
%(Others put their funding somewhere around here.)
}

% Bibliography
%\bibliography{Gauss-Euler}

%\end{document}

%%%%%%%%%%%%%%%%%%%%%%%%%%%%%%%%%%%%%%%%%%
%%%%%%%%%%%%%%%%%%%%%%%%%%%%%%%%%%%%%%%%%%
%%%%%%%%%%%%%%%%%%%%%%%%%%%%%%%%%%%%%%%%%%
%%%%%%%%%%%%%%%%%%%%%%%%%%%%%%%%%%%%%%%%%%
%%%%%%%%%%%%%%%%%%%%%%%%%%%%%%%%%%%%%%%%%%
%%%%%%%%%%%%                                               %%%%%%%%%%%%%%%
%%%%%%%%%%%%                       SI                     %%%%%%%%%%%%%%%
%%%%%%%%%%%%                                               %%%%%%%%%%%%%%%
%%%%%%%%%%%%%%%%%%%%%%%%%%%%%%%%%%%%%%%%%%
%%%%%%%%%%%%%%%%%%%%%%%%%%%%%%%%%%%%%%%%%%
%%%%%%%%%%%%%%%%%%%%%%%%%%%%%%%%%%%%%%%%%%
%%%%%%%%%%%%%%%%%%%%%%%%%%%%%%%%%%%%%%%%%%
%%%%%%%%%%%%%%%%%%%%%%%%%%%%%%%%%%%%%%%%%%

\newpage
\clearpage

\renewcommand{\thefigure}{S\arabic{figure}}
\setcounter{figure}{0} 
\renewcommand{\theequation}{S\arabic{equation}}
\setcounter{equation}{0}

%\begin{center}
\noindent\textbf{\LARGE Supplementary Information for \\} \\
%\noindent\textbf{Supplementary Information for \\
\noindent\textbf{\Large ``Mesoscale structure of wrinkle patterns and defect-proliferated liquid crystalline phases" \\} \\
%\end{center}
\noindent\textbf{\footnotesize Oleh Tovkach, Junbo Chen, Monica M. Ripp, Teng Zhang, Joseph D. Paulsen, Benny Davidovitch}
\vspace{0.75in}

In Secs.~1-6 %of this {\emph{Supplementary Information (SI)}} 
we provide relevant background on the model system used in our paper, and expand on various technical details of the theoretical developments that are described succinctly in the main text. In Sec.~7 we expand on some technical aspects of the experiments and simulations, as well as data analysis.

{\bf Equations that are introduced in the {\emph{SI}} are labeled as ``$S$[number]''. Equation labels that are not preceded by "S" refer to the main text.}

%\begin{abstract}

%\end{abstract}

%\keywords{wrinkles | thin sheets | periodic structures | metrology}

%\abbreviations{FvK, F\"{o}ppl-von K\'{a}rm\'{a}n; FT, Far from Threshold}

%\dropcap{W}rinkle patterns in compressed thin sheets are ubiquitous in nature and in technology, from the furrows on a forehead, to crinkly plant leaves, from ripples on plastic-wrapped objects to the protein film on milk. Perhaps the most elementary descriptor of these patterns is the wavelength of wrinkles. Understanding the dependence of the wavelength on the properties of the sheet and of the underlying liquid or elastic subphase allows us to exploit the wrinkle pattern as a tool to sculpt surface topography~\cite{Genzer06,Breid11}, to measure properties of the sheet~\cite{Stafford04,Huang07, Schroll13}, or to infer the forces applied to it~\cite{Burton97}. There is now a reasonable understanding of the wavelength of wrinkle patterns that are parallel, spatially uniform, and planar~\cite{Cerda03}. However, naturally occurring wrinkles typically do not satisfy one or more of these stipulations. In this article, we present a scheme that quantitatively explains the wrinkle wavelength beyond such idealized situations.
%%%%%%%%%%%%%%%%%%%%%%%%%%%%%%%%%%%%%%%%%%%%%%%%%%    
\section{Review of the Lam\'e problem}
\label{sec:Lame}
In this section we briefly review some essential features of the Lam\'e set-up, emphasizing key aspects related to tension field theory, far-from-threshold analysis, and the energetic hierarchy underlying the wrinkled state.  
%%%%%%%%%%%%%%%%%%%%%%%%%%%%%%%%%%%%%%%%%%%%%

\subsection{Axisymmetric state and tension field theory}
\label{sec:axi-tft}

The axisymmetric (unwrinkled) solution of the Lam\'e problem is described by a classical solution \cite{timoshenko70}.  Considering for simplicity $\Rin \ll \Rout$, the displacement field $\rmu_r(r)$, and the radial and hoop stress components are given by: 
\begin{gather}
\srr(r) = \gin \cdot [\frac{\Rin^2}{r^2} (1-\frac{1}{\tau}) + \frac{1}{\tau}]  \ \ ;  \ \ \sqq(r)
= \gin \cdot [\frac{\Rin^2}{r^2} (-1+\frac{1}{\tau}) + \frac{1}{\tau}]  \ , 
 \nonumber \\
\rmu_r(r)  =  \frac{\gin}{Y} \cdot [\frac{\Rin^2}{r} (-1+\frac{1}{\tau}) + \frac{r}{\tau}] 
\ ,  
\label{eq:sol-Lame}
\end{gather}  
and the elastic energy (including the work done on the sheet by the tensile boundary loads), 
$\tfrac{1}{2Y}\int rdrd\q \ \sigma_{ij}^2 -  2\pi \gin \cdot [ \tau^{-1} \Rout \rmu_r(\Rout) -  \Rin \rmu_r(\Rin) ]  $, is given by: 
\begin{equation}
U^{\text{(Lam\'e)}} (\tau) = - \pi \Rin^2\frac{\gin^2}{Y} \cdot \{ -1 + 2(\frac{1}{\tau} - 1)^2  +
\frac{\Rout^2}{\Rin^2}\frac{1}{\tau^2} \} \ . 
\label{eq:en-Lame}
\end{equation}  
(In the above expressions we use dimensional and dimensionless parameters as defined in the main text. Additionally, we simplified the expression by taking the Poisson's ratio $\Lambda = 0$, recalling that the Poisson's ratio does not affect the wrinkle pattern \cite{Davidovitch12,Taylor15}).   

For $\tau >2$, the axisymmetric state is characterized by hoop compression ($\sqq (r) <0$) in the zone $\Rin < r < L^{\text{(Lam\'e)}} = \sqrt{\tau -1}\ \Rin$. 
Hence, if the sheet is sufficiently thin, %For sufficiently small $\epsilon$ ({\emph{i.e.}} small bending modulus), 
the axisymmetric state is unstable to the formation of radial wrinkles that relieve the compressive stress. More precisely, there is a ``threshold curve'',  
$\tau_w(\epsilon, \Bo) >2$, such that a system characterized by $\epsilon, \Bo$ and $\tau > \tau_w (\epsilon, \Bo)$ is unstable to the formation of wrinkles. Furthermore, %denoting $\ttau_c  =  \tau_c(\epsilon, \Bo)-2$, 
one finds that $\tau_w(\epsilon, \Bo) \to 2$ as $\epsilon \to 0$, reflecting the familiar fact that the buckling threshold vanishes with the sheet's thickness. 
\\

Tension field theory (TFT) describes the stress field and radial displacement %in the wrinkles state far from threshold conditions, namely,  
for a given $\tau >2$ in the {\emph{singular}} limit $\epsilon = 0$, at which bending rigidity vanishes and the sheet cannot support any compressive stress.  As was described in detail in Refs. \cite{Davidovitch11,Davidovitch12}, in this limit, the stress field is purely tensile, so that the sheet ``splits'' into two parts: 

{\emph{(a)}} a wrinkled zone, $\Rin<r<L(\tau) = \tfrac{1}{2}\tau \Rin $, in which the hoop compression ``collapses'', and the radial stress is tensile, fully determined by (radial) force balance and the tensile load exerted on the inner edge: 
\begin{equation}
\srr(r) = \gin \Rin/r  \ \ ; \ \  \sqq(r) = 0  \   \ \ (\Rin<r<L(\tau))  \ . 
\label{eq:sol-TFT}
\end{equation} 
{\emph{(b)}} an unwrinkled zone, $L(\tau)<r<\Rout$, at which both radial and hoop stress components are tensile ({\emph{i.e.}} positive), and are described by Eq.~(\ref{eq:sol-Lame}) upon substituting $\Rin \to L(\tau)$, $\gin \to 2\gout = 2 \gin/\tau $, and $\tau \to 2$. The radial displacement in the wrinkled zone is given by: 
\begin{equation}
 \rmu_r(r)  =\frac{\gin}{Y} \cdot [ (-1 + \log(\frac{r}{L(\tau)}))\Rin+ \frac{2}{\tau}L(\tau) ] < 0 \   \ \ (\Rin<r<L(\tau)) 
\label{eq:ur-TFT}
\end{equation}  
which underlies a ``slaving condition'' on the wrinkled state, namely, that the fractional arclength ``wasted'' by wrinkly undulations in a latitude of radius $r$ in the wrinkled zone is: 
\begin{equation}
\Phi^2(r)  = -\rmu_r(r)/r  \ ,  \ \ (\Rin<r<L(\tau))
\label{eq:phi-TFT}
\end{equation}    
(Eq. 5 of main text). 
If one totally neglects the energetic cost of bending ({\emph{i.e.}} imagining a hypothetic sheet with no bending rigidity), the above tension field solution describes the stress and radial displacement in a wrinkled state at mechanical equilibrium, which is energetically favorable in comparison to the axisymmetric (compressed, unwrinkled) state. Namely, considering the energy stored in the (purely tensile) strain and the work done by the boundary loads, one finds the energy:  
\begin{equation} 
U_{\rm dom} (\tau) = - \pi \Rin^2\frac{\gout^2}{Y} \tau^2\cdot \left[ -\frac{1}{2} + \log(\frac{\tau}{2}) 
%-1 + 2(\frac{1}{\tau} - 1)^2  
+ \frac{\Rout^2}{\Rin^2}\frac{1}{\tau^2} \right] \ . 
%\label{eq:en-Lame}
\label{eq:en-TFT}
\end{equation} 
The corresponding expression in the main text, Eq. 6 (from which we omitted the uniform contribution $\propto \Rout^2\gout^2/Y$, that scales with the sheet's size) is valid also for nonzero Poisson ratio. 
%ignores terms that vanish in the limit of infinite size sheet, $\Rout/\Rin \to \infty$).}  

%{\color{red} Mention difference from Eq. in text. Include effect of Rin Rout, also-- change gin to gout and insert tau2}

According to TFT, for any $\tau>2$ the dominant contribution to the energy~(\ref{eq:en-TFT}) is lower than its counterpart, $U^{\text{(Lam\'e)}} (\tau)$ (Eq.~\ref{eq:en-Lame}), associated with the axisymmetric (compressed, unwrinkled) state.   
%%%%%%%%%%%%%%%%%%%%%%%%%%%%%%%%
 
 \subsection{Far-from-threshold analysis}     
 The basic premise of a ``far-from-threshold'' (FT) approach is an implementation of tension field theory to study the mechanical equilibrium of highly bendable sheets, namely, physical sheets with very small, but nevertheless nonzero bending modulus. More precisely, defining 
 \begin{equation}
 \ttau_w(\Bo,\epsilon) = \tau_w(\Bo,\epsilon) -2 \ , 
 \label{eq:ttau}
 \end{equation}
 one may distinguish, for given values of $\Bo$ and $0<\epsilon \ll 1$, between the two parameter regimes: $\tau-2 \gtrsim \ttau_w(\Bo,\epsilon)$  
 %= \ttau_c(\Bo,\epsilon)$ 
 and $\tau - 2 \gg \ttau_w(\Bo,\epsilon)$. In the former (near threshold) regime, mechanical equilibrium may be found by standard post-buckling analysis, namely, an expansion around the axisymmetric, unwrinkled state, Eqs.~(\ref{eq:sol-Lame},\ref{eq:en-Lame}), in which the small parameter is the wrinkle amplitude. In the latter, ``far-from-threshold'' regime, the expansion is around the tension field solution, Eqs.~(\ref{eq:sol-TFT}-\ref{eq:en-TFT}), and the small parameter of the expansion is the inverse-bendability, $\epsilon$. Since $\ttau_w(\Bo, \epsilon) \to 0$ as $\epsilon \to 0$, the far-from-threshold approach is prevalent for highly bendable sheets, and needs to be applied in fact for every $\tau >2$, %but at a 
except for an extremely narrow sliver in the parameter space.
 
% \subsubsection{How small should $\epsilon$ be ?}
\subsubsection{Identifying the far-from-threshold regime}

Our current study addresses the FT parameter regime and implements the corresponding theoretical approach, hence it is important to characterize the threshold curve, 
$\tau_w(\Bo, \epsilon) = 2+ \ttau_w(\Bo, \epsilon) $. Previous studies focused on the case $\Bo=0$ ({\emph{i.e.}} no liquid sub-phase), where scaling arguments show that $\ttau_w(\Bo=0, \epsilon) \sim \epsilon^{1/4}$ \cite{Davidovitch11,Davidovitch12}. Here, our primary interest is in $\Bo \gg 1$, such that the wrinkle wavelength is governed by the substrate stiffness (associated with the liquid g.p.e.). One may estimate the wrinkling threshold by comparing -- for a given set of $\tau, \Bo, \epsilon$ -- the {\emph{residual}} hoop compression in the wrinkled state with the {\emph{bare}} compressive stress, namely, the ``would-be'' compressive hoop stress, had wrinkles not been formed. For $\Bo \gg 1$, the former is just $|\sqq| \approx 2\sqrt{\Ksub B}$,  
%(since we consider here he regime $\Bo \gg 1$), 
whereas the latter -- evaluated at $r=\Rin$, can be estimated as $\approx \gin (\tau/2-1)$ \cite{Davidovitch11}. Hence, the bare compression exceeds the residual level if $\tau  > 2 + \ttau_w$, where: 
\begin{equation}
\ttau_w(\Bo,\epsilon) \approx 4 \sqrt{\Bo\cdot\epsilon}  \ . 
\label{eq:thresh}
\end{equation}        
%Since in our study we focus on $\Bo \gg 1$, and 
Since in our analytical study we implement the FT methodology for the parameter regime $\Bo \gg 1$, whereas the value of %range of 
$\tau$ %values 
explored in our experiments and simulations are no larger than 10, %rather limited ($2.5<\tau<10$), 
Eq.~(\ref{eq:thresh}) implies that a meaningful comparison between theoretical predictions and simulations/experiments requires the values of $\epsilon$ to be sufficiently small, namely, $\epsilon^{-1} \gg \Bo \gg 1$. (Specific numerical values are given in Sec. 7).      

 %In order to explore the FT parameter regime with   
 %Explain the threshold condition, and criterion for when FT analysis is valid (residual compressive stress  $\ll $ bare compressive stress).  
%show why it requires $\epsilon \ll 1$ for $Bo \ll 1$ and $Bo \epsilon \ll 1$ for $Bo \gg 1$.   
 
 \subsubsection{Energetic hierarchy}
%The principle that underlies the far-from-threshold analysis is {\emph{energetic hierarchy}}. 

Since for given values of the parameters 
$\tau,\Bo$, and $\epsilon=0$, the energy minimum is given by %$U_{\rm dom} (\tau)$ (
Eq.~(\ref{eq:en-TFT}), %one expects that for a small but finite $\epsilon$, 
the energy minimum for a small but finite $\epsilon$ can be expressed through an expansion around this limit value:
\begin{equation}
U(\tau, \Bo, \epsilon) = U_{\rm dom}(\tau) + U_{\rm sub-dom}(\tau,\Bo, \epsilon)  %+ \cdots  \ , 
\label{eq:en-hierarchy}
\end{equation} 
such that $U_{\rm sub-dom}(\tau,\Bo, \epsilon) \to 0$ as $\epsilon \to 0$, and the energy minimum approaches the tension field value, $U(\tau, \Bo, \epsilon)  \to U_{\rm dom}(\tau)$ in this limit. 
%(where the power $1/2$ is required by considerations of force and energy balance, see \cite{Davidovitch11,Davidovitch12,Bella14}). 

Notwithstanding the fact that the value of the sub-dominant energy is negligible in comparison to the dominant term, its mere existence 
%its minimization 
is crucial for understanding the nature of the wrinkle pattern. The reason is that there are many wrinkle states that are consistent with the tension field limit, and it is thus the minimization of the sub-dominant energy, which determines the value of $U_{\rm sub-dom}(\tau,\Bo, \epsilon)$, %that 
and thereby selects the physical state. %, which minimizes the whole energy. 
The primary purpose of our article is to characterize the {\emph{meso-scale structure}} of the energy-minimizing wrinkle state that corresponds to the value of $U_{\rm sub-dom}(\tau,\Bo, \epsilon)$, where $\tau >2, \Bo \gg 1$ and $\Bo\cdot \epsilon \ll 1$. 
Our theory does this 
%A basic tenet of our theory is 
%that this is % the minimum of $U_{\rm sub-dom}(\tau,\Bo, \epsilon)$
%obtained 
by minimizing a Ginzburg-Landau energy functional $U_m + U_{\psi} + U_{\rm nonlin}$ (Eqs. 10,15,16), which expresses the FvK energy of a wrinkle pattern in the vicinity of $r=r_a$ through an %described by an 
ansatz: %of the form 
\begin{equation}
\zeta(r,\theta) = {\rm Re} [\Psi(r,\theta)e^{im_a \theta}] 
\label{eq:ansatz}
\end{equation} 
%{\color{red} Fix this. Either change to maintext notation, or introduce $\theta_a$} 
(Eq. 12), subject to the condition that the underlying 
stress field and radial displacement (thereby the fractional wasted arclength of latitude $\Phi^2(r)$) are given by the tension field solution (Eqs.~\ref{eq:sol-TFT},\ref{eq:ur-TFT},\ref{eq:phi-TFT}). The degrees of freedom of this functional are the coarse-grained fields $\Psi (r,\theta)$ 
%(whose spatial variation is over scales much larger than the wrinkle wavelength),  
and the number $m_a = 2\pi r_a /\lambda(r_a)$,  
%$\lambda (r,\theta_a)$ 
%= 2\pi r/m(r,\theta_a)$, %\hat{n}(\bfx)$, 
that describe, respectively, the spatial variation of the magnitude and phase of the (complex) wrinkle amplitude, and the 
periodic rapid oscillations of the pattern. (In principle, the wrinkle wavelength may vary also slowly in the azimuthal direction, namely, 
$\partial_\theta \lambda(r_a,\theta) \neq 0$, but we do not delve here into this possibility).      

\subsubsection{The bending-substrate energy functional} % ($U_m$)}
The term $U_m$ (Eq. 10) %of the energy functional, 
describes the energy cost due to bending and substrate deformation. 
In contrast to the other contributions to the coarse-grained energy functional, $U_\psi$ and $U_{\rm nonlin}$ (Eqs. 15-16), $U_m$ is a quadratic functional of the magnitude of the complex amplitude, $|\Psi|$, which does not involve its gradients. The physical meaning of this mathematical difference is that $U_m$ can be obtained by considering the wrinkled zone as a continuous set of decoupled {\emph{elastica}}-like rings (obtained by ignoring the term $\srr \partial_r^2$ in Eq. 7), whereas 
$U_\psi$ and $U_{\rm nonlin}$ account for energetic penalties beyond this simplistic picture that emerge when one accounts for the non-zero value of $\srr$. We show in Sec.~3 that all three terms of the functional can be obtained by computing the FvK and substrate deformation energy for the ansatz (\ref{eq:ansatz}). However, it is useful to show how $U_m$ emerges naturally if one adopts the simplified picture of decoupled {\emph{elastica}}-like rings.   

Consider then a continuous set of decoupled {\emph{elastica}}-like rings of radii $\Rin<r <L(\tau)=\Rin\tau/2$, attached to a substrate of stiffness 
$\Ksub$, each of them subjected to confinement $\Phi^2(r)$ implied by the tension field solution (Eq.~\ref{eq:phi-TFT}). Furthermore, assume that the shape is given by the variable-separated ansatz ({\emph{i.e.}} Eq. 8, or equivalently, Eq.~\ref{eq:ansatz} with $\partial_{\theta}\Psi  = 0$), such that the contribution to the bending energy due to $\partial_{\theta}\Psi  \neq 0$ is incorporated into the energy functional $U_{\psi}$. %(see next section). 
For each such ring, the mechanical equilibrium states are given (assuming $\Phi^2(r) \ll 1$) by sinusoidal undulations (Eq. 8), parameterized by the ``wavenumber'' $m$, and the energetic cost of bending and substrate deformation is then: %for each ring (of infinitesimal width $dr$) is then: 
\begin{multline}
U_m  = \int dr \int_0^{2\pi}r d\theta \  |\Psi |^2 \left( \frac{1}{2}B (\frac{1}{r^2} \frac{\partial^2 \cos(m\theta)}{\partial \theta^2})^2  +    
\frac{1}{2}K \cos^2(m\theta)  
 \right) \\
 %\frac{1}{4} dr \int_0^{2\pi} r d\theta \  |\Psi |^2 \left( \frac{1}{2}B (\frac{1}{r^2}+    \frac{1}{2}K  \right)
 =
\int dr \int_0^{2\pi} r d\theta \ \Phi^2 r^2 \left(B r^{-4} m^2  + K m^{-2} \right) \ , 
\label{eq:integra-Um}
\end{multline}
where we used Eq. 5, and retained the integral over $\theta$ for convenience. Minimizing over $m$, one obtains $\barm(r)  = r(\Ksub/B)^{1/4}$ (Eq. 11). Expanding around this minimal value and exploiting once again Eq. 5, we obtain the energy functional $U_m \{\Psi, m \}$ given by Eq. 10. 

%%%%%%%%%%%%%%%%%%%%%%%%%%%%%%%%%%%%%%%%
 
\subsubsection{Beyond a ``local-$\lambda$ law''}
Previous studies of the far-from-threshold regime have focused on the ``micro-structure'' of the wrinkle pattern, namely, the energetically-favorable average wavelength, $\blambda(r)$, obtained upon ignoring the fact that the parameter $m$ in the variable-separated ansatz (Eq. 8) is an integer, whose spatial variation requires the presence of localized defects.    
%rather than on meso-scale structure, such as the amplitude modulation, which is the primary focus of the current work. 
%For that purpose, previous works typically assumed the variable-separated ansatz (Eq. ? of the main text), and incorporated 
In that approach, the effect of transverse tension $\srr$ (as well as curvature-induced effects in cases where wrinkles form on a non-planar background), is incorporated by replacing the substrate stiffness $\Ksub$ with the stiffness of an ``effective substrate'', $\Keff \approx \Ksub + K_{\rm tens} + K_{\rm curv}$, where $K_{\rm tens} = \srr [\Phi'(r)/\Phi(r)]^2$ accounts for the effect of transverse tension, $\srr$, on suppressing the wrinkle amplitude (and $K_{\rm curv}$ accounts for a similar amplitude-suppression effect due to transverse curvature, which is not relevant for the Lam\'e problem) \cite{Paulsen16}. In this framework, the sub-dominant energy 
$U_{\rm sub-dom}(\tau,\Bo, \epsilon)$
in Eq.~(\ref{eq:en-hierarchy}) is approximated by minimizing a renormalized version of the energy functional $U_m$, with $\Ksub \to \Keff$. Such a balance between a locally-determined effective stiffness and bending energy has been called the ``local-$\lambda$ law''.

While such an approach is useful for describing the small deviation ($\sim \Bo^{-1}$) of the average value of $\blambda$ from $2\pi (B/\Ksub)^{1/4}$ due to the presence of radial tension, it overlooks the strong deviation of the amplitude from the form assumed by the variable-separated anstaz (Eq. 8), and specifically the azimuthal modulations of the amplitude, and the crucial distinction between defect-rich and defect-free patterns (Figs. 2,4 of the main text). Our modified ansatz, Eq.~(\ref{eq:ansatz}), through which the wrinkle amplitude $\Psi(r,\theta)$ becomes a free variable rather than a ``slaved'' one, along with describing the sub-dominant energy functional explicitly ($U_m + U_\psi + U_{\rm nonlin}$) rather than merely renormalizing $U_m$ by $\Ksub \to \Keff$,    
%$U_m$ (which accounts for substrate energy) and $U_\psi$ (governed by radial tension), 
constitute a minimal model for describing the {\emph{meso-scale}} structure of the wrinkle pattern, which emerges from the conflict between an energetically-favorable micro-scale ({\emph{i.e.}}  the wavelength $\blambda$) and an incompatible macro-scale geometry ({\emph{i.e.}} the director $\hat{n} = \hat{\theta}$).

%%%%%%%%%%%%%%%%%%%%%%

\subsection{Capillary-induced tension on a floating sheet}
Our experimental system consists of a liquid drop placed at the center of a large, ultrathin floating sheet (see main text). While this ``drop on sheet'' problem has been a subject of intensive studies in recent years \cite{Huang07,Schroll13,Toga13}, the aspect that is most relevant for the current paper is the wrinkle pattern observed in the {\emph{exterior}} of the sheet-drop contact line, namely, where the sheet is flat except azimuthal undulations ({\emph{i.e}} $\int d\theta \zeta(r,\theta) \approx 0$). As was shown in previous studies \cite{Schroll13,Davidovitch18a} this part of the sheet can be thought of as a Lam\'e set-up, where the tension $\gamma_{\rm out} = \gamma_{lv}$ is given by the liquid-vapor surface tension that pulls on the sheet's edge and $\gamma_{\rm in} \propto \gamma_{lv}^{2/3}Y^{1/3}$ is induced by the capillary tension of the drop that pulls the sheet inward at the contact line. (In fact, the coefficient in the last scaling relation is not merely a constant, but rather 
$|\log(\gamma_{lv}/Y)|^{-1}$, see {\emph{e.g.}} Fig. 4 of Ref.~\cite{Davidovitch18a}). The crucial point is that since $Y \gg \gamma_{lv}$, the ratio $\tau = \gamma_{\rm in} /\gamma_{\rm  out} \sim (Y/\gamma_{lv})^{1/3}$ is generally much larger than the critical value $\tau_w= 2 + \ttau_w(\Bo,\epsilon)$, above which the axisymmetric state of the sheet consists of a hoop-compressed zone, and the system is therefore unstable to the emergence of radial wrinkles.

%FT seeks to describe the mechanical equilibrium for values of $\tau>2$, for which  $(\tau - 2)/ \ttau_c(\Bo, \epsilon) \gg 1$, through an expansion around the the tension field solution      

%of the dimensionless parameters, $\tau, \Bo$, and $\epsilon$,      

%More precisely, considering a triplet of dimensionless parameters, $\tau, \epsilon, \Bo$, such that $\tau -2 \gg \ttau_c (\epsilon, \Bo)$, the    

%Review Lam\'e TFT -- explain matching conditions and $U_{dom}(\tau)$. Explain energetic preference to Lam\'e solution $U_{Lame}(\tau)$.  

%\subsection{The subtle effect of tension-induced stiffness}   

%%%%%%%%%%%%%%%%%%%%%%%%%%%%%%%%%%%%

\section{Multi-scale analysis of the FvK equation}   
%{\color{red} this needs quite a bit of revision. Specifically, explain what is $\Boa, \theta_a$, etc. Also - explain here how we know that $\theta_a$ is $\Boa^{-1/2}$.}
In order to perform a multi-scale analysis of the resonant perturbation problem (Eq. 7) in a narrow, defect-free annulus of radius $r_a$, we start by introducing a local Bond number: 
\begin{equation}
\Boa = \Ksub r_a^2 /\srr(r_a)  \ ,  
\label{eq:Boa-0}
\end{equation}   
where $\srr(r_a)$ is given by the TFT solution~(\ref{eq:sol-TFT}),  
and assume an expansion of the deflection $\zeta(r,\theta)$ in powers of $\Boa^{-b}$, where $b>0$, of the form:   
%
%Implementing multi-scale analysis for the resonant perturbation problem (Eq. 7), we 
%consider an expansion in $\Boa^{-b}$ of the form: 
\begin{equation}
%\zeta (r,\theta) = \zeta_0(r,\theta_r,\theta_a) + \Boa^{-1/2} \zeta_1(r,\theta_r,\theta_a) +  \Boa^{-1} \zeta_2(r,\theta_r,\theta_a) + \cdots \ ,
\zeta (r,\theta) = \zeta_0(r,\theta_r,\theta_a) + \Boa^{-b} \zeta_1(r,\theta_r,\theta_a) +  \Boa^{-2b} \zeta_2(r,\theta_r,\theta_a) + \cdots \ ,
\label{eq:expansion-2} 
\end{equation} 
where: 
\begin{gather}
\theta_r = \theta \ \ ; \ \  
%\theta_a = \frac{m_a}{\sqrt{\Boa}} \theta \nonumber \\
\theta_a = m_a \Boa^{-b} \ \theta  \ , 
%\nonumber \\
%\Boa = \Ksub r_a^2 /\srr(r_a) 
\label{eq:Boa}
\end{gather} 
and: 
\begin{equation}
\zeta_0(r,\theta_r,\theta_a)   = {\rm Re} [\Psi(r,\theta_a) e^{im_a\theta_r}]  = \frac{1}{2} 
[\Psi(r,\theta_a) e^{im_a\theta_r} + c.c.]
%\overline{\Psi} (r,\theta_a) \exp^{-im_a\theta_r} ] 
\ . 
\label{eq:zeta0}
\end{equation} 
\\

%\underline{Comment 1:} The reason that the expansion (\ref{eq:expansion-2}) is carried out through $\Boa^{-1/2}$ is clarified in the sequel. 

%\underline{Comment 2:} 
The expression~(\ref{eq:zeta0}) for the leading term in the expansion~(\ref{eq:expansion-2}) is a more sophisticated (but mathematically-equivalent) form of the generalized ansatz we introduced in the main text (Eq. 12), to assist us with the analysis. This is a standard technique in multi-scale analysis, whereby a single variable ($\theta$) is decomposed into two ``independent'' variables ($\theta_r$ and $\thetasl$) such that the ``rapid'' and ``slow'' variations of the function $\zeta(r,\theta)$ are demarcated and can be analyzed distinctly. %We define the (yet unknown) exponent $b$, such that the variation of the amplitude    
Using this approach, differentiation w.r.t. $\theta$ becomes:            
%We introduced explicitly the ``rapid'' coordinate $\theta_r =\theta$ to clarify the multi-scale technique, whereby a function of two independent variables ($r,\theta$) is analyzed as a function of three independent variables ($r,\theta_r,\theta_a$), such that: 
\begin{equation}
%\frac{\partial \zeta_0}{\partial \theta} = \frac{\partial \zeta_0}{\partial \theta_r} + \frac{m_a}{\sqrt{\Boa}} \frac{\partial \zeta_0}{\partial \theta_a}  = 
%\frac{1}{2} [im_a \Psi e^{im_a\theta_r} + \frac{m_a}{\sqrt{\Boa}}  
%\frac{\partial \Psi}{\partial \theta_a} e^{im_a\theta_r}  + c.c.]
\frac{\partial \zeta_0}{\partial \theta} = \frac{\partial \zeta_0}{\partial \theta_r} + {m_a}{\Boa}^{\!\!\!\!-b} \frac{\partial \zeta_0}{\partial \theta_a}  = 
\frac{1}{2} [im_a \Psi e^{im_a\theta_r} + {m_a}{\Boa}^{\!\!\!\!-b}  
\frac{\partial \Psi}{\partial \theta_a} e^{im_a\theta_r}  + c.c.]
%im_a (\Psi \exp^{im_a\theta_r}-  \overline{\Psi}\exp^{-im_a\theta_r})    +  \frac{m_a}{\sqrt{\Boa}}  
%(\frac{\partial \Psi}{\partial \theta_a} \exp^{im_a\theta_r} + \frac{\partial \overline{\Psi}}{\partial \theta_a}  \exp^{-im_a\theta_r})  
\ . 
\end{equation}    

A crucial point to understand about the multi-scale expansion technique is that we {\emph{do not}} seek here to solve explicitly for the next terms  in the expansion (\ref{eq:expansion-2}), {\emph{i.e.}} $\zeta_1,\zeta_2, etc.$ but merely to find a {\emph{solvability condition}} that guarantees the existence of such an expansion. This solvability condition turns out to be precisely the amplitude equation (Eq. 13). 

To see this, let us re-express the operator ${\cal L}_0$ (Eq. 7) through the new coordinates: 
\begin{gather}
 {\cal L}_0 = {\cal L}_0^r +  {\cal L}_0^a \nonumber \\
{\cal L}_0^r = B\frac{1}{r^4}\frac{\partial^4}{\partial \theta_r^4} -\stt \frac{1}{r^2}\frac{\partial^2}{\partial \theta_r^2} + \Ksub  \nonumber \\
{\cal L}_0^a = B\frac{1}{r^4}(\frac{\partial^4}{\partial \theta^4}  - \frac{\partial^4}{\partial \theta_r^4})   -\stt \frac{1}{r^2}
(\frac{\partial^2}{\partial \theta^2} -\frac{\partial^2}{\partial \theta_r^2})   \ , 
\label{eq:operator-expand}
\end{gather}    
and use this expression for an expansion in powers of $\Boa^{-b}$ of the equation: 
\begin{equation}
 [{\cal L}_0 + {\cal L}_1]\zeta = 0  \ , 
 %(r, \theta) = 0  \ , 
 \label{eq:333}
\end{equation}
where ${\cal L}_1 = -\srr \partial^2_r$ (Eq. 7), and the dependence of $\zeta$ on the three coordinates ($r,\theta_r,\theta_a$) is given by Eq.~(\ref{eq:expansion-2}). %, through the small parameter, $\Boa \ll 1$, we obtain 

At $O(\Boa^0)$ we obtain: 
\begin{equation}
{\cal L}_0^r [\zeta_0]  =  \Psi(r,\theta_a) {\cal L}_0^r [e^{im_a\theta_r}]  +c.c =0  \ , 
%[\Psi(r,\theta_a) e^{im_a\theta_r}] \Psi(r,\theta_a) =  
\end{equation}  
which is satisfied for {\emph{any}} $\Psi(r,\theta_a)$, and the amplitude-wavelength slaving condition in Eq. 5 only imposes a global constraint, which does not exclude $\partial_{\theta_a} \Psi(r,\theta_a) \neq 0$ !

At the next orders in $\Boa^{-b}$, substitution of Eqs.~(\ref{eq:expansion-2},\ref{eq:operator-expand}) into Eq.~(\ref{eq:333}), yields a series of non-homogenous equations for $\zeta_1,\zeta_2, \cdots$ of the form: 
\begin{gather}
{\cal L}_0^r [\zeta_1] = (nonhom)_1  \nonumber \\
{\cal L}_0^r [\zeta_2] = (nonhom)_2  \nonumber  \ , 
\end{gather}
and so on, where the terms ``$(nonhom)_i$'' originate from operating with ${\cal L}_0^a + {\cal L}_1$ on the lower order terms ($\zeta_j$ with $0\leq j<i$) in the expansion~(\ref{eq:expansion-2}). In order for these equations to be solvable, the ``Fredholm alternative'' \cite{HabermanBook}  implies the orthogonality of the 
$(nonhom)_i$ terms and %to %are orthogonal 
%to 
the zero modes ({\emph{i.e.}} functions within the kernel) of the adjoint of ${\cal L}_0^r$, namely:
\begin{equation}
\int r dr \int_0^{2\pi}d\theta_r [e^{im_a\theta_r}  +c.c] \cdot  (nonhom)_i   = 0  \  . 
\label{eq:integrate-1}
\end{equation}   
%(this is known as the ``Fredholm alternative''). 
Note that it is sufficient to perform the integral over an infinitesimally narrow annulus, centered at $r=r_a$ (and furthermore -- it is sufficient to limit the integration to an azimuthal sector, $0<\theta_r<2\pi/m_a$, namely, the period of the rapid wrinkly undulations). This allows us to avoid the radial integration and replace $r \to r_a$ in the above integral. Focusing on the leading non-homogenous term, $(nonhom)_1$, we find that:
%{\color{red} (here need to fix numbers)}  
\begin{gather}
(nonhom)_1 = - ({\cal L}_0^a + {\cal L}_1) \{\zeta_0\}   \nonumber  \\
= - \frac{1}{2} \left\{ 
B\frac{m_a^4}{r_a^4}   \left( 4 \Boa^{-b} (-i) (\frac{\partial\Psi}{\partial \theta_a} e^{im_a\theta_r} -c.c.) - 6 
\Boa^{-2b} (\frac{\partial^2\Psi}{\partial \theta_a^2} e^{im_a\theta_r} + c.c.)  \right. \right. \nonumber  \\
\left. \left. + 4\Boa^{-3b} (+i) (\frac{\partial^3\Psi}{\partial \theta_a^3} e^{im_a\theta_r} -c.c.) + \Boa^{-4b} (\frac{\partial^4\Psi}{\partial \theta_a^4} e^{im_a\theta_r} + c.c.) \right) \right. \nonumber \\
\left. 
%+ 2 
-
\sqq \frac{m_a^2}{r_a^2}  \left( \Boa^{-b} (+i) (\frac{\partial\Psi}{\partial \theta_a} e^{im_a\theta_r} -c.c.) +  
\Boa^{-2b} (\frac{\partial^2\Psi}{\partial \theta_a^2} e^{im_a\theta_r} + c.c.) \right) 
-\srr(r_a)  (\frac{\partial^2\Psi}{\partial r^2} e^{im_a\theta_r} + c.c.)  \right\} \ . 
\label{eq:solvability-1}
\end{gather} 
%This allows us to consider both $r$ and $\theta_a$ that appear in as constants $r$ and $\theta_a$   terms are    
Substituting the RHS of Eq.~(\ref{eq:solvability-1}) in Eq.~(\ref{eq:integrate-1}), we find that the terms proportional to $\frac{\partial\Psi}{\partial \theta_a}$ 
%(and also $\frac{\partial^3\Psi}{\partial \theta_a^3}$) 
vanish upon integration over $\thetara$, so that for $\Boa \gg 1$ %the leading order contribution to %$\Boa^{-1}$, 
Eq.~(\ref{eq:integrate-1}) becomes: 
%are of order $O(\Boa^{-2b})$   
\begin{equation}
4 \Boa^{-2b}\frac{1}{{r_a^2}} \frac{\partial^2\Psi}{\partial \theta_a^2}  + \Boa^{-1} \frac{\partial^2\Psi}{\partial r^2}  = 0  \ , 
\label{eq:nnn}
\end{equation}
%so that to leading order in $\Boa^{-1}$ the solvability condition involves only even derivatives of $\Psi$ w.r.t. $\theta_a$ and $r$. 
where we have used $\sqq = -2\sqrt{B\Ksub}$ and $m_a = r_a(\Ksub/B)^{1/4}$ (Eqs. 9,11), and the definition (\ref{eq:Boa}) of $\Boa$. \\
%and considering the terms in the expansion ({\emph{i.e.}} $\sim \Boa^{-1}$), 
%we obtain Eq. 13.     

Inspection of Eq.~(\ref{eq:nnn}) reveals that the only nontrivial value of the exponent $b$, for which the two terms remain comparable in the limit $\Boa \to \infty$, is $b=1/2$. More precisely, if $b>1/2$, then the solvability condition~(\ref{eq:integrate-1}) becomes $\frac{\partial^2\Psi}{\partial r^2} = 0$, similarly to the monochromatic ansatz, and consequently the hurdles that motivated us to introduce a generalized ansatz with azimuthal dependence of the amplitude are not resolved (see paragraph prior to Eq. 12 in the main text). Conversely, if $0<b <1/2$, then the solvability condition~(\ref{eq:integrate-1}) becomes $ \frac{\partial^2\Psi}{\partial \theta_a^2} = 0$, which is independent of the actual resonant perturbation ({\emph{i.e.}} the operator ${\cal L}_1$ 
in Eq. 7) that forced us to replace the monochromatic ansatz (Eq. 8, for which $\Psi$ is independent of $\theta$) with a generalized one that allows a $\theta$-dependent amplitude (Eq. 12).

Using Eqs.~(\ref{eq:Boa-0},\ref{eq:Boa}) with $b=1/2$ to transform back from $\thetasl$ to $\theta$, Eq.~(\ref{eq:nnn}) reduces to Eq. 13 of the main text.
%*********

%Comments: 

%1) can replace $r \to r_a$, etc. Also, no need to consider corrections to of $\Bo$ to $\stt$.      
%%%%%%%%%%%%%%%%%%%%%%%%%%%%%%%%%%%%%%%%%%

%%%%%%%%%%%%%%%%%%%%%%%%%%%%%%%%%%%%%%%%%%
\section{Coarse-grained energy}
%%%%%%%%%%%%%%%%%%%%%%%%%%%%%%%%%%%%%%%%%%
In this section, we show how the coarse-grained energy functional, $U_m + U_\psi + U_{\rm nonlin}$, whose minimization underlies %which we use to approximate 
our approximation of the 
sub-dominant energy, $U_{\rm sub-dom}$ in Eq.~(\ref{eq:en-hierarchy}) and thereby the meso-scale structure of the wrinkle pattern, is obtained from coarse-graining the FvK energy of the sheet and the substrate energy. We start by briefly reviewing the various parts of the FvK and substrate energies, and then show how their expansion around the tension field limit, with respect to the small parameters $\epsilon \ll 1$ and $\Boa^{-1} \ll 1$, where the ansatz (\ref{eq:zeta0}) is assumed, yields the energy functional, $U_m + U_\psi + U_{\rm nonlin}$. 

\subsection{The FvK and substrate energies}      
%A crucial point here is that the strain, and consequently the 
The in-plane strain, and consequently the FvK energy of an elastic sheet, cannot be expressed through the out-of-plane displacement component $\zeta(r,\theta)$ alone. Instead it requires also the in-plane components. % (this underlies the non-linear nature of FvK equations).  
As long as the exerted tensile strain is small ($\gin/Y \ll 1$), the slopes remain small even in the fully-developed wrinkled state ({\emph{i.e.}} $|\nabla \zeta| \ll 1$), such that the displacement can be expressed through the Monge parameterization: 
%using cylindrical coordinates: % \cite{notations-u}:
\begin{equation}
\mathbf{u}(r,\q) = \ru_r(r,\q)\mathbf{\hat r} +  \ru_\q(r,\q)\boldsymbol{\hat\q} +  \zeta(r,\q) \boldsymbol{\hat z} \   , 
\label{eq:displacement}
\end{equation}
and the components of the strain tensor, $\boldsymbol\varepsilon$, are: 
%\subsubsection*{Strain and stress}
%The strain tensor $\boldsymbol\varepsilon$ is given by: 
\begin{subequations} \label{eq:strains}
\begin{gather}
\varepsilon_{rr} = \partial_r \ru_r + \tfrac{1}{2} (\partial_r\zeta)^2\ , \label{eq:strain-radial-1} \\
\varepsilon_{\q\q} = \tfrac{1}{r} \partial_\q \ru_\q + \tfrac1r \ru_r + \tfrac{1}{2r^2} (\partial_\q \zeta)^2 \ ,\label{eq:hoopstrain}\\
\varepsilon_{r\theta} = \varepsilon_{\theta r} = \tfrac12 \left( \tfrac1r \partial_\q \ru_r +  \partial_r \ru_\q + \tfrac1r \partial_r\zeta\partial_\q \zeta\right)\ .  \label{eq:strain-shear-1}
\end{gather}
\end{subequations}
%The stress in the sheet is given by the Hookean relationship \cite{LL86,Timoshenko,MansfieldBook}: 
%so that this relationship is linear (Hookean response). 
%Denoting the strain tensor $\boldsymbol\epsilon$, and using a polar coordinate system {\emph{with its origin at the center of the film}}, which is the natural choice for our study, this Hookean relationship becomes \cite{LL86}: 
%\begin{subequations} \label{eq:stresses}
%\begin{gather}
%\sigma_{rr} = \frac{Y}{1-\Lambda^2} \left(\varepsilon_{rr} + \Lambda \varepsilon_{\q\q} \right)\ \label{eq:stresses-radial} , \\
%\sigma_{\q\q} = \frac{Y}{1-\Lambda^2} \left( \varepsilon_{\q\q} + \Lambda \varepsilon_{rr} \right)\ , \\
%\sigma_{r\theta} = \frac{Y}{1+\Lambda} \varepsilon_{r\theta} \ , 
%\end{gather}
%\end{subequations}
%where $Y=E_ft$ is the stretching modulus and 
%where $\Lambda$ the Poisson ratio of the sheet \cite{Comment-Disp-Strain-Stress}. %Again, we emphasize that Eqs.~(\ref{eq:strains}) describe the geometric (strain-displacement) connection only for $|\nabla \ru| \ll 1$ (which neccesarily implies $|\epsilon_{ij}| \ll 1$), where the Hookean (stress-strain) response only requires $|\epsilon_{ij}| \ll 1$ (which may be satisfied even in situation where the gradient $|\nabla \ru|$ is not small and the geometric connection, Eq.~(\ref{eq:strains}), is not valid.   
%\subsubsection*{Curvature}
Furthermore, %anticipating the shape $\zeta(r,\theta)$ to be characterized by small slopes (even in the wrinkled zones), 
the curvature tensor $\kappa_{ij}$ can be approximated as: 
\begin{equation}
\kappa_{rr} = \partial^2_{rr}\zeta \ \ ; \ \ \kappa_{\theta\theta}  = \tfrac{1}{r} \partial_r \zeta + \tfrac{1}{r^2} \partial^2_{\theta\theta} \zeta \ \ ; \ \ \kappa_{r\theta} = 2\partial^2_{r\theta} \zeta 
\label{eq:disp-curvature}
\end{equation} 
The energy, which we denote by a capital $U$, is conveniently expressed through its areal density (denoted by italic lower case, $u={U}/{\rm area}$): 
\begin{subequations} \label{eq:FvKenergy}
\begin{equation}
%u = u_{\rm FvK} + u_{\rm Win} \ , 
u = \uFvK + \usub \ , 
\label{eq:EnWinModel}
\end{equation}
where $u_{\rm FvK} = u_{\rm strain} + u_{\rm bend}$ , and: %(Eq.~\ref{eq:FvKschem}), and:   
\begin{gather}
\usub  = \frac{\Ksub}{2}  \zeta^2  \ ,  \nonumber  \\ % (\zeta-r^2/2R)^2 \ ,  \nonumber  \\
u_{\rm strain} = \frac{1}{2} \sigma_{ij} \varepsilon_{ij}  \ \ ; \ \ 
u_{\rm bend} = \frac{B}{2} Tr\boldsymbol({\bf \kappa})^2 \approx \frac{B}{2} \left(\frac{1}{r^2} \frac{\partial^2\zeta}{\partial \theta^2}\right)^2 
\label{eq:EnWindefine} \nonumber \\
{\rm where}: \sigma_{ij} = Y\varepsilon_{ij} 
\end{gather}
\end{subequations}
%(Eq.~5 of main text). 
%In Eq.~(\ref{eq:EnWindefine}) 
(As mentioned already, for simplicity of the presentation we take in the above equations a zero Poisson ratio, noting that the analysis for a non-zero Poisson ratio requires more bookkeeping but does not affect any of the results that are relevant for our study.)    
%Above we we once again employed the inequality, $W\ll R$, approximating the rest (spherical) state of the substrate as $\zeta_{sph}(r) \approx r^2/R$. 
Furthermore, since the curvature is governed by %bending energy is barely affected by 
%we will see later that the bending energy is  governed by 
the wrinkly undulations (rather than the radial variation of their amplitude), we retained in the above expression only the contribution from the component $\kappa_{\theta\theta}$ to the bending energy.  \\
%profile rapid, azimuthal undulations of the shape, and hence can be safely approximated as: $u_{\rm bend} \approx \tfrac{B}{2}  r^{-2}(\partial_{\q\q} \zeta)^2$ ({\emph{i.e.}} variation of the shape along the radial direction make negligible contribution to the energy).     

%%%%%%%%%%%%%%%%%%%%%%%%%
%\subsection{The dominant energy(or: the tension field limit for displacement and strain)} 
\subsection{Principles of the expansion}
The tension field theory solution, described in Sec.\ref{sec:Lame}, yields the asymptotic state of the radial displacement, and correspondingly the confinement function $\Phi^2(r)$ that determines the fraction of the latitudinal arclength wasted by undulations in the limit $\epsilon \to 0$ (for any $\Boa$). These limit values are obtained by minimizing the energy $\Ustrain$ along with the work done by the tensile loads $\gin, \gout=\gin/\tau$, that pull on the edges, subject to the compression-free condition, $\sqq \geq 0$. In that (singular) limit, $\Uwin = \Ubend = 0$.   

Denoting by $\Delta \Ustrain$ the difference between the strain energy in the actual wrinkled state and the tension field limit, our purpose here is to express the various contributions to the sub-dominant energy $U_{\rm sub-dom}(\tau,\Bo, \epsilon)$ in Eq.~(\ref{eq:en-hierarchy}) by evaluating     
 %Our purpose here is to expand the energies 
$\Ubend + \Uwin + \Delta \Ustrain$, for $0<\epsilon \ll 1$ and $\Bo \gg 1$. We do this   
by assuming the ansatz (\ref{eq:zeta0})
and re-organizing the various terms in 
$\Ubend + \Uwin + \Delta \Ustrain$ into an energy functional, $U_m +  U_\psi + U_{\rm nonlin}$, whose field variables are  
 the wavelength $\lambda$ and complex amplitude $\Psi$, and whose parameters are derived from the known features of the tension field limit.        \\

%At a technical level, the basic difference between our approach and previous developments of the far-from-threshold approach is based on the nature of the  ansatz for $\zeta$. Those previous works assumed the variable-separated ansatz (?),  which does capture only an average micro-scale, namely $\blambda (r)$, but fails to describe how the actual wavenumber $m(r) $ varies between integer values, and consequently overlooks the meso-scale structure associated with this spatial variation. As we discussed above (Sec. ??,), in that approach the subdominant energy is  found by minimizing the functional $U_m$ only (albeit with a renormalized version of the effective substrate stiffness $\Keff$), and the wrinkle amplitude is totally determined by the wavelength through the slaving condition (?), and is not considered as a free field variable.  In contrast, the current work, aimed at revealing the meso-scale structure underlying wrinkle patterns, employs the more sophisticated ansatz (? of the main text), for which the field variables of the functional underlying the sub-domiant energy are the wavelength $\lambda$ {\emph{and}} the complex amplitude $\Psi$. \\

\underline{\emph{Small parameters of the expansion:}} %Before dwelling on technical details of the calculation, l
Let us mention that in addition to the control parameters: 
\begin{equation}
\epsilon \ll 1 \ \ {\rm and} \ \  \Boa \gg 1 \ ,
\end{equation}
 there are yet  two other emergent ratios, whose assumed smallness is employed %``hidden'' small parameters that we employ 
 in our analysis, namely:
 \begin{equation}
\ellt/r_a \ll 1  \ \ {\rm and} \ \ \elln/r_a \ll 1 \ , 
\label{eq:cond-3}
\end{equation}  
where $\ellt$ and $\elln$ define the sizes of a defect-free zone in the hoop and radial directions, respectively (see Fig. 3 of main text). The condition $\ellt \ll r_a$ is guaranteed self-consistently by the results of our analysis (Eq. 17 or 18), whereas $\elln \ll r_a$ follows from the assumed low cost of defect energy, assisted by amplitude suppression (see the paragraph following Eq. 13 in the main text). We stress that a complete coarse-grained theory of the defect-proliferated wrinkled state should yield also an actual prediction for $\elln$, in terms of the various control parameters ({\emph{e.g.}} $\gin/Y, \tau, \epsilon, \Boa$), in an analogous manner to %similarly to 
the prediction of $\ellt$ (Eq. 18). We suspect that such a prediction requires one to consider yet higher orders in the expansion (in $\Boa^{-1}$, $\epsilon$, and possibly parameters that involve the stretching modulus, {\emph{e.g.}} $\gin/Y$), and furthermore, must take into consideration explicitly the actual energetic cost of defects. The current version of our theory does not provide tools to evaluate the actual energetic cost of defects, and therefore it 
%being limited to leading order in that expansion, 
is capable of predicting only $\ellt$ but falls short from providing an analogous prediction for $\elln$.     
     
%%%%%%%%%%%%%%%%%%%%%%%%%%%%%%%%%%
\subsection{Strain and energy densities}    
We consider then a deformation of the sheet in the zone $\Rin<r<L(\tau)$, whose out-of-plane component is given by the ansatz %(\ref{eq:ansatz}). 
(\ref{eq:zeta0}). 
In  addition to the value of $L(\tau)$, tension field theory implies two other conditions. First, the radial displacement is given, up to corrections that vanish as $\epsilon \to 0$, by Eq.~(\ref{eq:ur-TFT}), and the arc-length wasted by $\zeta(r,\theta)$ is given by $\Phi^2(r)$, Eq.~(\ref{eq:phi-TFT}), up to corrections that vanish as $\epsilon \to 0$. In the following, we provide the corresponding expressions for the strain components, $\varepsilon_{rr}$ and $\varepsilon_{\q\q}$ (in the Lam\'e problem, the shear strain, $\varepsilon_{r\q}$ contributes to the energy only at a higher order in $\epsilon$ \cite{Davidovitch12}):     
\begin{gather}
\varepsilon_{rr} %(r,\thetara,\thetasl) 
%=  \frac{\partial \rmur}{\partial r} +\frac{1}{2}\left(\frac{\partial\zeta}{\partial r}\right)^2 \nonumber \\
%\nonumber \\
%\frac{\partial \bar{\rm u}_r}{\partial r} +\frac{1}{2}\left(\frac{\bzeta}{\partial r}\right)^2  
%-\frac{1}{2}\frac{\partial^2\bzeta}{\partial r^2}(\Psi(r,\theta) e^{i m_0 \theta} +c.c.)
%+\frac{1}{8}( \frac{\partial\Psi}{\partial r})^2 (e^{2i m_0 \theta} +c.c.)   \nonumber \\
= \frac{1}{Y}\srr (r) \  + \ \frac{1}{4}\cdot \left|\frac{\partial\Psi}{\partial r}\right|^2  
%\nonumber \\ 
+ \frac{1}{8} [( \frac{\partial\Psi}{\partial r})^2 e^{2i m_a \thetara} +c.c.]
%-\frac{1}{2}\frac{\partial^2\bzeta}{\partial r^2}[\Psi e^{i \thetara} +c.c.]
 \ ,
%
%+\frac{\partial\zeta_0}{\partial r}\frac{\partial \Psi}{\partial r}\cos\left[m_0\theta\right] +\frac{1}{2}\left(\frac{\partial f}{\partial r}\right)^2 \cos^2\left[m\theta\right]\right\}  \\
%= \sigma_{rr}^0 -Yf\frac{\partial^2\bar{\zeta}}{\partial r^2}\cos\left[m\theta\right] +\frac{Y}{2}\left(\frac{\partial f}{\partial r}\right)^2 \cos^2\left[m\theta\right]
\label{eq:res-epsrr-nn}
\end{gather}
and
\begin{gather}
\varepsilon_{\theta\theta}= %(r,\thetasl) =
%\frac{1}{Y}\bsqq  \  +  \ 
\left( 
 \frac{m_a^2}{4 r^2}
 \left|\Psi\right|^2  -\Phi^2 (r)\right) \ \nonumber  \\
 + 
 \    \frac{m_a^2}{4r^2} 
 %\deltaz^2 
 \Boa^{-1} \left|\frac{\partial\Psi}{\partial\thetasl}\right|^2  \ + \ 
   \frac{m_a^2}{4 r^2} \left( i
   %\deltaz 
   \Boa^{-1/2} 
   \Psi\frac{\partial\Psi^*}{\partial\thetasl}  +c.c. \right)  \  
 \label{eq:res-epstt-nn}  
 \ ,   
 %\\ + i \delta \frac{m^2}{4r^2} \left(\Psi\frac{\partial\Psi^*}{\partial\thetasl} - \Psi^*\frac{\partial\Psi}{\partial\thetasl} \right) \nonumber 
\end{gather}
where $\srr(r)$ is given by Eq.~(\ref{eq:sol-TFT}). 
Note that, in contrast to the last term in %terms in the second line of 
Eq.~(\ref{eq:res-epsrr-nn}), the expression for $\varepsilon_{\q\q}$ does not include terms that oscillate ``rapidly'', such as $\frac{m_a^2}{4 r^2} |\Psi|^2 e^{2im_a\thetara}$. The reason is that such terms can be made to cancel out by properly adjusting the azimuthal displacement, ${\rm u}_{\theta}$ (which contributes to the hoop strain through $r^{-1}\partial_\theta {\rm u}_{\theta}$, see Eq.~(\ref{eq:hoopstrain}). This observation was already noted in previous developments of the FT expansion \cite{Davidovitch12,Taffetani17,Davidovitch19}).

Using %Eqs.~(\ref{eq:energy-tot-0}-\ref{U_win}) 
Eq.~(\ref{eq:FvKenergy}) 
and the above expressions for the displacement~(\ref{eq:zeta0})
and strain components~(\ref{eq:res-epsrr-nn},\ref{eq:res-epstt-nn}), we can compute a coarse-grained version of the energy by integrating over the rapid variable, $\thetara$. The outcome of this calculation is three energy densities 
$\tilde{u}_{\rm strain}, \tilde{u}_{\rm bend},\tilde{u}_{\rm subst}$, such that $\tilde{u}_{(\cdot)} (r,\theta)\cdot r \Delta r \Delta\theta $ is the corresponding energy in a small annular zone of opening angle $\Delta\theta \approx 2\pi /m_a$ and a small radial width, $\Delta r \ll \ellt$,  
around a point $(r,\theta)$ in a defect-free annulus. Below we give a succinct version of these energy densities, omitting various terms which are negligible in comparison with other terms that appear elsewhere %explicitly 
in the following expressions, 
%({\emph{i.e.}} corresponding to higher orders in $\deltaz$ or $\barmz^{-1}$), 
such that ignoring them does not entail further constraints on energy minimization.

 %\noindent
%$\bullet$ 
The coarse-grained energy density $\tustrain$ is:  
%such that $\tPsi \sim O(1)$. %, helping us to recognize the order of various contributions to energy.
\begin{gather}
\tustrain  
%\frac{Y}{2} \frac{2\pi}{m_0}  \int_{0}^{2\pi} d\thetara  \ \varepsilon_{rr}(r,\thetara,\thetasl)^2 \nonumber \\
%=\  \frac{1}{2}\frac{\bsrr ^2}{Y}  \ +\  \frac{1}{4}\bsrr\left|\frac{\partial\Psi}{\partial r}\right|^2 \ +\  \frac{1}{4}\frac{Y}{R^2}\left|\Psi\right|^2 \ + \  \frac{3}{16}Y\left|\frac{\partial\Psi}{\partial r}\right|^4 \nonumber \\
%=\  \frac{1}{2Y}(\srr ^2 + \bsqq ^2)
=\  \frac{1}{2Y}\srr ^2 
%\frac{\bsrr ^2}{Y}  
%+  \frac{1}{2}\frac{\bsqq^2}{Y}  
%\ + \  \frac{Y}{2}\left(\left|\frac{r_0}{r}\tPsi\right|^2  -\bPhi^2 (r)\right)^2 
\nonumber \\
%\ + \ \frac{1}{m_0^2}\frac{Y}{R^2}r_0^2\left|\tPsi\right|^2 
%+  \  \frac{1}{m_0^2} r_0^2 \bsrr\left|\frac{\partial\tPsi}{\partial r}\right|^2 
+  \  \frac{1}{4} \srr\left|\frac{\partial\Psi}{\partial r}\right|^2 
\ + \ 
%\deltaz^2  \bsqq  \left|\frac{r_0}{r}\frac{\partial \tPsi}{\partial \thetasl}\right|^2 \nonumber \\
%\Boa^{-1}  \frac{1}{4} \sqq  m_a^2 \left|\frac{1}{r}\frac{\partial \Psi}{\partial \thetasl}\right|^2 \nonumber \\
\Boa^{-1}  \frac{1}{4} \sqq  m_a^2 \left|\frac{1}{r_a}\frac{\partial \Psi}{\partial \thetasl}\right|^2 \nonumber \\
%\ + \ \frac{Y}{2} \{\left|\frac{r_0}{r}\tPsi\right|^2 \!-\! \bPhi^2 (r)\}^2  \ +\   
\ + \ \frac{Y}{2} \{\left|\frac{m_a}{2r}\Psi\right|^2 \!-\! \Phi^2 (r)\}^2  \ +\   
%Y\deltaz^2 (\frac{r_0}{r})^2 {\rm {Im}} \{\tPsi\frac{\partial\tPsi^*}{\partial\thetasl} \}^2
%\frac{1}{16}Y\Boa^{-1} m_a^4 \frac{1}{r_a^2 r^2} {\rm {Im}} \{\Psi\frac{\partial\Psi^*}{\partial\thetasl} \}^2
\frac{1}{16}Y\Boa^{-1} m_a^4 \frac{1}{r_a^4} {\rm {Im}} \{\Psi\frac{\partial\Psi^*}{\partial\thetasl} \}^2 \ . 
% 
%\frac{Y}{2} \left(\{\left|\frac{r_0}{r}\tPsi\right|^2  -\bPhi^2 (r)\}^2 + 2  \deltaz^2 (\frac{r_0}{r})^2 {\rm {Im}} \{\tPsi\frac{\partial\tPsi^*}{\partial\thetasl} \}^2 \right)
%\nonumber \\
%+ \ Y \cdot \deltaz^2 \cdot (\frac{r_0}{r})^2 \cdot {\rm {Im}} \left[\tPsi\frac{\partial\tPsi^*}{\partial\thetasl} \right]^2
%\ + \  \frac{3}{m_0^4}Y r_0^4 \left|\frac{\partial\tPsi}{\partial r}\right|^4  \ . 
 \label{eq:c-g-urad-0-main}
\end{gather}   
%$\bullet$ 
The coarse-grained energy density $\tubend$ is: 
\begin{gather}  
\tubend  = 
%B \frac{m_0^2r_0^2}{r^4}  \left|\tPsi\right|^2   \nonumber  \\ +  \ 
\frac{1}{4}B \frac{m_a^4}{r^4}  \left|\Psi\right|^2   \nonumber  \\ +  \ 
%B\frac{m_0^2r_0^2}{r^4}\cdot \deltaz^2\cdot   \left(4\left|\frac{\partial\tPsi}{\partial\thetasl}\right|^2    -2 {\rm Re} \left[\tPsi \frac{\partial^2\tPsi}{\partial\thetasl^2}\right]   
%\frac{1}{4}B\frac{m_a^4}{r^4}\cdot \Boa^{-1}\cdot   \left(4\left|\frac{\partial\Psi}{\partial\thetasl}\right|^2    -2 {\rm Re} \left[\Psi \frac{\partial^2\Psi}{\partial\thetasl^2}\right] \right)
\frac{1}{4}B\frac{m_a^4}{r_a^4}\cdot \Boa^{-1}\cdot   \left(4\left|\frac{\partial\Psi}{\partial\thetasl}\right|^2    -2 {\rm Re} \left[\Psi \frac{\partial^2\Psi}{\partial\thetasl^2}\right] 
\right) \ . 
\label{eq:c-g-ubend-main}
\end{gather}
%$\bullet$ 
The coarse-grained energy $\tusub$ is:
\begin{equation}
\tusub =  \frac{1}{4}{\Ksub}\left|\Psi\right|^2 \ .  
\label{eq:c-g-usub-main}
\end{equation}
\\
(Since our analysis is based on the assumption that the ansatz (\ref{eq:zeta0}) is valid in a defect-free zone 
%We note that since our analys
%these coarse-grained energy densities are about to be integrated out over zones 
whose sizes are $\elln, \ellt \ll r_a$, we simplified the above expressions by replacing $\frac{1}{r}\partial_{\thetasl} \to 
\frac{1}{r_a}\partial_{\thetasl}$.)
\\
% in the various 
%appearances of $r$ in denominators.  

Let us inspect now the various terms in the above equations:

$\bullet$ The first line of Eq.~(\ref{eq:c-g-urad-0-main}) is already accounted for by tension field theory, and it is thus part of the tension field energy (\ref{eq:en-TFT}), along with the strain energy in the unwrinkled portion of the sheet and the work of the tensile boundary loads. Hence, this part of the energy is included in the term $U_{\rm dom}$ in Eq.~(\ref{eq:en-hierarchy}), and does not contribute to the sub-dominant energy. 

$\bullet$ Eq.~(\ref{eq:c-g-usub-main}), along with the first line of Eq.~(\ref{eq:c-g-ubend-main}), are quadratic in $|\Psi|$ and do not involve any gradients. Together, they form the functional $U_m$ (Eq. 10). %? of the main text). 

$\bullet$ The terms in the last line of Eq.~(\ref{eq:c-g-urad-0-main}) underlie the functional $U_{\rm nonlin}$ (Eq. 15). Note that in order to get this we switched back $\thetasl \to m_a\theta/\sqrt{\Boa}$.       

$\bullet$ Finally, the terms in the second lines of Eqs.~(\ref{eq:c-g-urad-0-main},\ref{eq:c-g-ubend-main}) underlie the functional $U_\psi$ (Eq. 16). To see this, note that: 

{\emph{(a)}} we once again switched back $\thetasl \to m_a\theta/\sqrt{\Boa}$. 

{\emph{(b)}} 
Since we consider $\Boa \gg 1$, we approximated $\sqq \approx -2\sqrt{B\Ksub}$ and $m_a \approx r_a (\Ksub/B)^{1/4}$.

{\emph{(c)}} Anticipating that $\Psi$ can be approximated by a real function (see main text and Sec. 4), the parenthetical term in the second line of Eq.~(\ref{eq:c-g-ubend-main}) transforms as follows: %is approximated by:  
$ 4(\frac{\partial\Psi}{\partial\thetasl})^2    -2 \Psi \frac{\partial^2\Psi}{\partial\thetasl^2}  = 6(\frac{\partial\Psi}{\partial\thetasl})^2 -2 \frac{\partial}{\partial\thetasl}(\Psi\frac{\partial \Psi}{\partial \thetasl})$. 

{\emph{(d)}}  The last term in the above expression is an exact derivative (in $\thetasl$) and thus amounts only to a boundary term, which we ignore (assuming that it vanishes upon integration over $\thetasl$) . 

%{\color{red} 
%\subsection{The amplitude equation as an Euler-Lagange equation of the coarse-grained energy functional}%}
%{\color{red}  do this}
%\newpage
     
%which is suggested -- as we saw above -- by an asymptotic expansion of the FvK equation in the limit $\Bo \gg 1$. Here we show how the three components of the energy functional ..  , are obtained from substituting this ansatz in the FvK and substrate energies . .  

%\subsection{Expansion in $\epsilon$ and $\Bo^{-1}$}      

\section{Minimizing the coarse-grained energy %subjected amplitude equation 
subject to near-inextensibility constraints}
% and the Euler-Lagrange equation of the coarse-grained energy functional}

When addressing purely developable deformations of thin solid bodies ({\emph{i.e.}} that do not affect the midplane's Gaussian curvature), the Euler {\emph{elastica}} principle implies that it is possible to consider a purely inextensible limit, where the energy and residual stress do not depend on the stretching modulus, but only on the bending modulus and the energy associated with boundary tensile loads and deformation of a substrate. An analogous idea underlies the variable-separated ansatz (Eq. 8), assumed in previous studies of radial wrinkle patterns in general, and the Lam\'e problem in particular. Although a pattern of wrinkles with radial orientation is clearly a non-developable deformation of the naturally-planar sheet, the variable-separated ansatz is constructed such that all hoops remain inextensible, namely, satisfying Eq. 5 for every $\Rin<r<L$.
 %(and a  non-zero, purely tensile strain, is fully accommodated by the radial component of the strain tensor). 
 For such an ansatz, the residual compressive stress would have been $-2\sqrt{B \Ksub}$, independent of the stretching modulus, exactly as for a uni-axially compressed {\emph{elastica}} supported on a substrate of stiffness $\Ksub$. 
%However, although the variable-separated ansatz is sufficient to predict the average wavelength, $\blambda(r)$, it does not account for the meso-scale modulation of the wrinkle amplitude and the pattern of defects enabled by it. 

\subsection{The amplitude equation as an Euler-Lagrange equation of the coarse-grained energy functional}%}   
  
As we noted in the main text (paragraph preceding Eq. 17), a $\theta$-dependent amplitude $\Psi(r,\theta)$  
%amplitude-modulated ansatz (\ref{eq:zeta0}) that we study here 
is not compatible with the strict amplitude-wavelength slaving condition (Eq. 5) for every $r$, amounting to violation of the perfect hoop inextensibility; consequently, the residual compression is larger than $2\sqrt{B \Ksub}$ and depends on the stretching modulus $Y$. However, assuming the energetic cost of defects in amplitude-suppressed zones is negligible, such that the size of defect-free zones may be very small (Eq.~\ref{eq:cond-3}), we can overcome this difficulty by imposing suitable conditions on the amplitude. As we noted in the main text (paragraph following Eq. 16), a key element in our analysis is the energetic hierarchy associated with the three components of the coarse-grained energy: $U_m$ (Eq. 10), $\Unonlin$ (Eq. 15), and $U_\psi$ (Eq. 16). Hence, our approach to minimize the sum $\Unonlin + U_m + U_\psi$ is to consider the first two terms as implying suitable constraints on the minimization of $U_\psi$. More precisely, we assume that the actual values of the energies $\Unonlin$ and $U_m$ in each defect-free zone are much smaller than the value of $U_\psi$. Focusing on  a single defect-free zone, this approach allows us to consider $m$ and $\srr$, $\sqq$ Eq.~16 as fixed parameters, given by Eqs.~4,9,11, since
any deviations from these values will incur an explicit energetic cost of $\Unonlin$ or $U_m$. Consequently, the only degree of freedom in the energy functional $U_\psi$ (Eq.~16) is the complex amplitude $\Psi(r,\theta)$. A standard variational calculus yields a Laplace-like equation (Eq. 13).     

As we noted in the main text, in order to determine the energetically-favorable solution of Eq. 13 we must recall the assumed negligibility of $\Unonlin$, on which consideration we elaborate in the next two subsections.            
         
%Nevertheless, we show below that Eq. 17 (or its more general version, Eq. 18) together with Eq. (\ref{eq:cond-3}) 
%guarantee that the residual hoop stress is only slightly perturbed from this value, such that the energetic hierarchy underlying tension field theory 
%(\ref{eq:en-hierarchy}) remains valid. A second implication of %the necessity to maintain 
%this energetic hierarchy is the persistence of nearly-radial orientation of wrinkles within defect-free zones and the consequent disruption of local smectic order in them.  

\subsection{A nearly radial orientation of wrinkles}   
%As was noted in the main text, 
The level of local smectic order in defect-free zones is governed by a competition between two parts of the coarse-grained energy functional. The first is $U_m$ (Eq. 10), which favors a local smectic order with fixed spacing of wrinkles ({\emph{i.e.}} $\lambda = \barlambda$) and hence would favor a local, nonuniform rotation of the bent director $\naux(\bfx) \to \hat{n}(\bfx)$, such that $\nabla \times \ \hat{n}(\bfx) \approx 0$. Since variation of the the director is related to the phase of the complex amplitude, this means that in each defect-free zone $U_m$ favors: $\partial_\theta \arg\Psi = \barm(r)-m_a$. The competing term is the second part in the integrand comprising $U_{\rm nonlin}$ (Eq. 15), which penalizes deviations from radial orientation, hence favors $\partial_\theta \arg\Psi  = 0$. In order to determine which of the two dominates, we can evaluate the difference in the values taken by each of these two terms upon substituting these two values of $\partial_\theta \arg\Psi$ in the respective functionals.  

{\emph{(a)}} For $U_m$, substituting $\partial_\theta \arg\Psi = \barm(r)-m_a$ (which is equivalent to $m = \barm(r)$) yields a zero value for the integral term in Eq. 10. Substituting $\partial_\theta \arg\Psi = 0$ and Taylor-expanding $(m_a- \barm(r))^2$ with $m_a = \barm(r_a)$, around $r=r_a$ (analogously to (\ref{eq:Taylor-1})), we find the energy density $\sim \sqrt{B\Ksub} \Phi^2(r_a) (\elln/r_a)^2$ (where the Taylor expansion procedure allows us to use $m \approx m_a$ and $|\Psi(r)| \approx |\Psi(r_a)|$).      

{\emph{(b)}} For $U_{\rm nonlin}$, substituting $\partial_\theta \arg\Psi = 0$  ({\emph{i.e.}} perfectly radial wrinkles) yields a zero value upon integrating the second integrand in Eq. 15. Substituting $\partial_\theta \arg\Psi = \barm(r)-m_a$ and using similarly a Taylor expansion around $r=r_a$, we find the energy density $\sim Y \Phi^4(r_a) (\elln/r_a)^2 \sim \gin \Phi^2(r_a) (\elln/r_a)^2 $ (up to a $\tau$-dependent prefactor, where we used Eq. 5).  

The ratio between these respective energetic costs is thus $\sqrt{B\Ksub}/\gin \sim  \sqrt{\Bo\cdot \epsilon}$. However, as was shown in Sec. 1.B.2, this is precisely the parameter whose decreasing value (and asymptotic vanishing) characterizes the transition from the instability-threshold condition to the far-from-threshold regime. Hence, as the wrinkle pattern becomes far from threshold ({\emph{e.g.}} decreasing the bending modulus  
%sheet thickness 
while keeping all other physical parameters are fixed), we expect the energetic cost of $U_{\rm nonlin}$ to be the dominant among the two terms, and hence the wrinkles in the defect-free zones to become more radially-oriented. 

In order to understand the implication of this finding on the meso-scale structure of the pattern we recall that our analysis assumes a defect-proliferated state and specifically (\ref{eq:cond-3}), such that the actual number of wrinkles in a defect-free zone is consistent with the ``correct'' number at the middle of the zone  
($\approx \ellt/\blambda$), and the deviation from a parallel array ({\emph{i.e.}} the relative angle between peaks) is constrained by the small ratio $\elln/r_a$.

\subsection{Weakening the amplitude-wavelength slaving constraint}
%As explained in the main text, t
The energetic cost of deviation from the amplitude-wavelength constraint (Eq.~5) is expressed by the first term in the integrand that comprises $U_{\rm nonlin}$ (Eq.~15). Considering a defect-free annular zone, $r \in (r_a-\elln, r_a + \elln)$, with $\elln \ll r_a$, where $m(r)=m_a = 2\pi r_a/\barlambda$, we can evaluate this term by considering a formal Taylor expansion around the value of the integrand at $r=r_a$: 
\begin{equation}
\left(\frac{m_a^2 |\Psi|^2}{4r^2} - \Phi^2(r)\right)^2 \approx  \left(A_0 + A_1 (\frac{r-r_a}{r_a}) + \frac{1}{2!} A_2(\frac{r-r_a}{r_a})^2 +\cdots \right)^2 \ ,  
\label{eq:Taylor-1} 
\end{equation}
where $A_i$ are the $i^{th}$ derivatives of $\left(\frac{m_a^2 \langle |\Psi|^2\rangle }{4r^2} - \Phi^2(r)\right)$ with respect to $r$ (evaluated a $r=r_a$) , and 
$\langle |\Psi|^2\rangle$ is averaged over azimuthal undulations of the amplitude. The energetic cost of deviations from the perfect slaving condition (Eq. 5) is determined by the first non-vanishing $A_i$. A nonzero $A_0$ would entail an energy areal density $\sim Y$, which will totally disrupt the TFT limit (Eq.~\ref{eq:en-TFT}). If $A_0=0$ but $A_1 \neq 0$, the energy density $\sim Y (\elld/r_a)^2$, and so on. Thus, minimization of the energetic penalty due to violation of the perfect slaving condition is tied to  
the azimuthal average of $\langle |\Psi|^2 \rangle$ and its radial derivatives.  

In order to proceed, we consider the Laplace-like equation (Eq. 13) in a narrow annulus, where natural basis functions are given by magnitude $c$ and wavelength $2\pi/d$: 
%spanned by the basis functions:
%that are periodic in $\theta$: 
\begin{equation}
\Psi(r,\theta) = c \ e^{\pm \sqrt{\tfrac{4|\sqq|}{\srr(r_a)}} d(r-r_a)} \cos(d \ r \ \theta)  \ .
\label{eq:pure-states}
\end{equation} 
Although any superposition of such states solves Eq. 13, the necessity to minimize the energy $U_{\Psi}$ (Eq. 16) motivates us to focus only on these ``minimal amplitude'' states, such that elimination of the two first coefficients $A_0=A_1=0$ in Eq.~(\ref{eq:Taylor-1}) provides two equations that determine $c,d$ in Eq.~(\ref{eq:pure-states}):   
\begin{gather}
\frac{{m_a}^2}{4}\langle \frac{\Psi^2}{r^2} \rangle_{r=r_a} = \Phi(r_a)^2 \ \ \Rightarrow \ \  
{c} = \sqrt{8}\frac{r_a}{m_a} \Phi(r_a) \label{eq:as}
\\
\frac{{m_a}^2}{4}\partial_r\langle \frac{\Psi^2}{r^2} \rangle_{r=r_a}  
\!=\! [\Phi(r_a)^2]' \ \ \Rightarrow \ \ 
%b \!=\! \frac{1}{2r_a^2}\frac{\Phi'(r_a)}{\Phi(r_a)} \ , 
d \!=\! \frac{1}{2}\left(\frac{1}{r_a} + |\frac{\Phi'(r_a)}{\Phi(r_a)}|\right) \sqrt{\frac{\srr(r_a)}{|\sqq|}}  \ . 
\label{eq:bs}
%a^2 =  .. 4 \rs^2 \ms^2 \Phi^2(\rs) \ \ , \\
%blabla 
\end{gather} 
The scale $\ellt$ characterizes the (azimuthal) distance between nearby zeros of the amplitude~(\ref{eq:pure-states}), namely: $\ellt \approx \pi/d$. Equation~(\ref{eq:bs}) together with the definition of $\ellBTs$ (Eq. 14), $\barlambda$ (Eq. 11), and $\sqq$ (Eq. 9) yield the  expression for $\ellt$ (Eq. 17) we report in the main text.  \\
 
 %The choice of the constants $\as,\bs$ in Eqs. 16-17 implies that the two leading terms in this Taylor expansion vanish ($A_0=A_1=0$), whereas $A_2$ may be an $O(1)$ term. Hence, the areal density of the energetic cost of the first term in $U_{\rm nonlin}$ is $\sim Y (\elld/r_a)^4 $. 
%(and proportional to the size $\sim \elld \ellt$ of a defect-free zone). This cost vanishes rapidly with the ratio $(\elld/r_a)$.    

Tension-field-based theory, which provides the basis for our analysis and whose assumed energetic hierarchy (\ref{eq:en-hierarchy}) ignores the explicit energetic cost of deviations from Eq. 5, remains valid if the energy $\Unonlin$ associated with the fact that $A_2 \neq 0$ in the Taylor expansion (\ref{eq:Taylor-1}) is small in comparison to the values of $U_m$ and 
$U_\psi$, whose simultaneous minimization comprises the sub-dominant energy $U_{\rm sub-dom}$ in (\ref{eq:en-hierarchy}). Specifically, considering $U_\psi$ (Eq. 16, whose minimization underlies the amplitude equation for $\Psi$, Eq. 13), whose characteristic value in a defect-free zone is characterized by energy density $\sim \srr$, we find that  
%(and proportional to the size $\sim \elld \ellt$). 
this inequality is valid as long as the density of defects is sufficiently large, such that $\elld/r_a \ll (\srr(r_a)/Y)^{1/4}$.       

%in that satisfy 
%Eqs.~(??) of the main text, the excess strain due to violation of the slaving condition (\ref{eq:slaving}) is $(\elld/r_0)^4$, and its energetic cost is negligible in comparison to the (minimum of) $U_{\Psi}$ in (\ref{eq:energy-2}) if $\elld/r_0 \ll (\srr*/Y)^{1/4}$. 

%must recall the ``slaving'' (\ref{eq:slaving}) imposed by the confinement. In contrast to the ansatz (\ref{eq:ansatz-mono}), this constraint cannot be satisfied throughout the annulus for the states~(\ref{eq:pure-states}). 
%However, the unavoidable violation of this constraint is minimized if: 
%since %recalling that 
%we consider narrow annuli, $\elld\!\ll\!\rs$, 
%we can find states that minimize {\emph{violation}} of this constraint, by choosing:
%However, recalling that we consider narrow annuli, $\ellt\!\ll\!\rs$, we can seek solutions that minimize violation of this constraint by requiring: 
%\begin{subequations}
%all $\theta$.}
%For $\as,\bs$ in %that satisfy 
%Eqs.~(\ref{eq:as},\ref{eq:bs}), 
%the excess strain due to violation of the slaving condition (\ref{eq:slaving}) is $(\elld/\rs)^4$, and its energetic cost is negligible in comparison to the (minimum of) $U_{\Psi}$ in (\ref{eq:energy-2}) if $\elld/\rs \ll (\srrs/Y)^{1/4}$. 

\section{Covariant formula for $\ell_{\parallel}$}
In deriving Eq.~17 of the main text we focused on the Lam\'e set-up, which is characterized by axial symmetry of the sheet and the tensile loads, hence it is not surprising that the corresponding prediction for $\ellt$ is explicitly dependent on the radial distance $r_a$ of a defect-free zone from the center of the sheet. One may wonder, however, whether a similar expression for $\ellt$ can be obtained for more generic problems, not necessarily characterized by such a global axial symmetry, in which the topography and confining forces give rise to confinement along an axis $\hat{n}(\bfx)$ that varies across the wrinkled zone. Assuming one finds (analytically or numerically) the TFT solution for such confinement problems, one may readily evaluate $\ellBTs$, $\barlambda$, and $\Phi(\bfx)$, but  there is no unique point in space that defines ``radial distance'', and therefore one must understand what length scale replaces $r_a$ in Eq. 18.   

If the director is bent in the vicinity of a point $\bfx$, namely $\nabla \times \naux (\bfx) \neq 0$, then a pattern of parallel, uniformly spaced wrinkles is not compatible with the confining conditions, and defect-rich patterns are likely to emerge, similarly to those in Fig.~2 of the main text. We note that our theoretical analysis is based on the existence  of a small, defect-free zone, in which the sheet undulates ``rapidly'' along a bent director ($\hat{\theta}$), such that the dependence on $r_a$ enters only through the azimuthal arclength formula, $ds = r_a d\theta$, which can be rewritten as: 
$ds  = |\nabla \times {\naux}(\bfx)|^{-1}  d\theta$.  Repeating our analysis with $r_a \to 
 |\nabla \times {\naux}(\bfx)|^{-1}$ yields Eq. 18 for confinement problems with locally bent director fields.       

\section{The transition curve $\Bo_c(\tau)$}
As mentioned in the main text, it is possible to obtain the scaling relation (Eq. 3) by considering the limit $\Bo \gg 1$ (while keeping 
$\sqrt{\Bo\cdot \epsilon} \ll 1$, such that our far-from-threshold expansion around the tension field limit is a valid approach), at which the transition from a defect-free to a defect-proliferated pattern is expected to occur when the wrinkled zone is limited to only a narrow annulus near the inner edge, namely, $0<\tau-2 \ll 1$. 
As we show below, in this limit it is possible to evaluate the difference in energies between these two states, using a similar approach to the one employed in Secs. 3-4  for evaluating energies in a single defect-free zone. 

{\emph{(a)}} Consider first the term $U_{\rm nonlin}$ of the energy (Eq. 15). For the defect-free state ({\emph{i.e.}} where the variable-separated ansatz is valid), both terms in the integrand of $U_{\rm nonlin}$ are identically zero. For the defect-proliferated state, we showed in Sec. 4 that both terms scale with the ratio $\elld/\Rin$ with powers that are $\geq 2$ . Hence upon integrating over the wrinkled annulus (where the number of defect-free zones is $\propto 1/\elld$), the overall value of $U_{\rm nonlin}$ for the defect-proliferated state scales at least as $\elld/\Rin$ (or as a higher power of this ratio). Consequently, if $\elld$ is sufficiently small in comparison to the radial width of the wrinkled annulus, namely, $\elld \ll \Rin(\tau/2-1)$, then we can ignore the effect of $U_{\rm nonlin}$ on the difference in energy between defect-free and defect-proliferated states.

{\emph{(b)}} Let us consider now the term $U_m$ of the energy (Eq. 10). For a defect-proliferated state (assuming ($\elld \ll \Rin (\tau/2 -1)$, such that $m \to m(r)\approx \barm(r)$), this term is close to its energy minimum, so that the integral term in Eq. 10 is nearly zero. In contrast, for a defect-free state, $m$ is a constant and minimization of the integral is realized by some $m^*  = \barm(r^*)$, where $\Rin<r^*<\Rin \tau/2$, 
so that  $(m - \barm(r))^2 = (m^* - \barm(r))^2 = \sqrt{\Ksub/B} (r-{r^*})^2$. Considering the rest of the terms in the integrand we obtain, upon using Eqs. 5,11, and expanding to leading order in $(\tau/2-1)$, a factor: $\propto (B\gin/Y \Rin^2) (\Rin\tau/2 - r)$. Hence, to leading order in $(\tau/2-1)$, the integral in Eq. 10 for the defect-free state scales as: $\propto  [\tfrac{\gin}{Y}\sqrt{B\Ksub} /\Rin^2] \int_{\Rin}^{\Rin \tau/2} (r-{r^*})^2 (\Rin \tau/2 -r) dr$. Since $\Rin<r^*<\Rin\tau/2$, the integral term is: $\propto \Rin^4 (\tau/2 -1)^4$, so that the difference in $U_m$ between the defect-free and defect-proliferated states is: 
$\Delta U_m \sim [\tfrac{\gin}{Y}\sqrt{B\Ksub} \Rin^2]  (\tau/2 - 1)^4$.   

{\emph{(c)}} Finally, let us consider the term $U_\psi$ of the energy (Eq. 16). Notably, the radial dependence of the excess latitudinal arclength ($\Phi^2(r)$, Eq. 5) implies that the first term in the integrand in Eq. 16 is nonzero for both defect-free and defect-proliferated states. Furthermore, previous studies (that addressed the case $\Bo=0$) found that the respective integral is diverging, and must be regularized by a boundary layer \cite{Davidovitch12} or another structure \cite{Bella14} at the vicinity of the wrinkle's foot ($r \to L = \Rin\tau/2$). Since this subtle regularization problem stems from only the near vicinity of the wrinkle's foot, it affects equally both defect-free and defect-proliferated states, and consequently we assume that the difference in $U_\psi$ between these two types is associated with the bulk of the wrinkled zone, $\Rin <r <\Rin\tau/2$, where $|\partial \Psi /\partial r|^2$ is finite (and can be estimated {\emph{e.g.}} by its value at the inner edge $r=\Rin$). 
Comparing Eqs. 8 and \ref{eq:as},\ref{eq:bs}, we note that the value of $|\partial \Psi /\partial r|^2$ is comparable in the two types of states (recall that we consider here $\elld \ll \Rin(\tau/2-1)$ so that the wrinkled annulus is densely populated by defects and correspondingly small defect-free zones). However, in contrast to the defect-proliferated state, the defect-free state (Eq. 8) does not have azimuthal modulations of the amplitude. Hence, since Eqs. 16  and \ref{eq:pure-states} show that the energetic cost of azimuthal modulations is a finite multiple of the cost of radial variation of the amplitude, we can estimate the difference $\Delta U_\psi$ through the radial width of the wrinkled annulus, $\Rin (\tau/2-1)$, multiplied by $|\partial \Psi /\partial r|^2$ evaluated at $r=\Rin$. 
Using Eq. 5 of the main text to evaluate $|\partial \Psi /\partial r|_{r=\Rin}$ we obtain, to leading order in $(\tau/2-1)$, 
$\Delta U_\psi \sim [\gin \tfrac{\gin}{Y}\sqrt{B/\Ksub}]$. %  (\tau/2 - 1)$. 

Comparing the above estimates for $\Delta U_m $ (which favors defect-proliferated states) and $\Delta U_\psi $ (which favors defect-free states), and recalling the definitions of the dimensionless parameters (Eq. 1), we obtain the scaling of the transition curve,  $\Bo_c(\tau) \sim (\tau -2)^{-4} $, whose scaling with $(\tau-2)$ is close to, but not identical to the scaling relation extracted from our data (Eq. 3 of the main text).
%in agreement with Eq. 3. 
As we noted in the main text, a likely reason for the deviation between the predicted and observed exponents is 
%the theoretical estimates  
%this agreement is remarkable given 
the fact that the data from experiments and simulations are taken at $\tau/2 -1  \sim O(1)$, whereas the above scaling analysis is focused on $\tau/2 -1 \ll 1$.

\section{Simulations and Experiments}

\subsection{Finite element simulations}\label{}

Finite element (FE) simulations were performed to simulate wrinkles in the Lam\'e setup with a wide range of values for $Bo \ ( 0.07 <{Bo} < 333)$, confinement ratio ($3.0 < \tau < 8.7$) and bendability ($5 \times 10^4 < \epsilon^{-1} < 3 \times 10^7$). Furthermore, in the large $\Bo$ regime, which is the primary focus of our study, we fixed the ratio $\sqrt{\Bo \cdot \epsilon} = 0.01$, to make sure the wrinkle pattern is well described by a far-from-threshold analysis around the tension field theory (see Sec. 1). 
  
The simulated films were made of elastic material with large Young's modulus and Poisson's ratio $\Lambda = 0.3$ to ensure the tension is much smaller than the in-plane stiffness ($\gamma_\text{in}/Y < 0.01$). The liquid surface tensions that pull on the inner and outer edges were modeled as tensile tractions along the radial direction of the undeformed thin film. 
%The traction directions do not follow shell rotation. 
The effect of gravity of the liquid substrate was modeled as a pressure on the sheet whose magnitude is proportional to the out-of-plane displacement of the element.

All the simulations were carried out with ABAQUS/Explicit. The 3-node linear shell elements (S3R) were used and geometric nonlinearity was taken into account. The pressure for modeling the liquid substrate was implemented through the user-defined subroutine for load distribution. A fine mesh was adopted at the inner edge with gradually increasing element sizes towards the outer edge. The FE nodes were randomly distributed to avoid any symmetry due to the discretized network. Large scale simulations with $10^6$ to $1.4 \times 10^7$ elements were required to capture the large number of wrinkles. A small initial pressure %($p = 10^{-7}$ to $10^{-9}$) 
was applied on the whole thin film to trigger wrinkles. The pressure then quickly decayed to zero and fictitious material damping was added to the model to help convergence to equilibrium. The simulation was then run dynamically until the out-of-plane deflection, 
%wrinkle magnitude profile 
$\zeta(r,\theta)$, converged to a stable state.

%\section*{Experiments}\label{}

%\subsection*{Film Preparation} 
%Films were made by spin-coating solutions of polystyrene ($M_\text{n} = 99$k, $M_\text{w} = 105.5$k, Polymer Source) in toluene ($99.9\%$, Fisher Scientific) onto glass substrates following Ref.~\cite{Huang07}. 
%A white-light interferometer (Filmetrics F3) was used to measure film thickness, which was uniform over each film to within $4\%$. 

\subsection{Wrinkle Analysis} 
Measurements %in Figs.~2e-h and 4e 
of the wrinkle number, $m(r)$, were obtained using a custom automated image analysis following Refs.~\cite{King12,Paulsen16}. 
%Image intensity was first averaged over small intervals along the radial coordinate to reduce noise. Then at each radius, the signal was filtered along $\theta$ to eliminate long-wavelength components due to uneven lighting, and an autocorrelation was performed, yielding a decaying oscillating signal. The wavelength of this oscillation was taken as the wrinkle wavelength. This analysis was performed on angular sectors free of material imperfections.
After an initial filtering step to reduce noise and lighting gradients, an autocorrelation of the intensity versus $\theta$ was performed at each radius within a region free of material imperfections, effectively averaging over many wrinkles. The wrinkle number was extracted from this oscillating signal. 
The same routine was used to analyze the wrinkle wavelength in the simulations by applying it to grayscale color maps of the out-of-plane deflection. 
%taken as the width of the first oscillation of this signal. 

\subsection{Determination of $\gamma_\text{out}$ and $\gamma_\text{in}$ in experiments}
%The confinement ratio, $\tau = \gin/\gout$, was extracted from the wrinkle length via a relation that was derived and validated previously for a finite-radius annulus in the far-from threshold regime \cite{Pineirua11,Taylor15}. 
%These measurements of $\tau$ were consistent with previous theory for the drop-on-sheet problem that addressed droplets smaller than the capillary-gravity length \cite{Schroll13}. 
The liquid-vapor surface tension, $\gamma_{lv}$, was measured with a Wilhelmy plate under typical experimental conditions, yielding $\gamma_\text{out}=\gamma_{lv}$. 
For each experimental image, the confinement ratio $\tau$ was deduced from the wrinkle length via a relation that was derived and validated previously for a finite annulus in the far-from-threshold regime \cite{Pineirua13,Taylor15}:
\begin{equation}
L = \frac{R_\text{out}}{\tau} \left( \frac{R_\text{out}}{R_\text{in}} - \sqrt{\frac{R_\text{out}^2}{R_\text{in}^2} - \tau^2} \right). 
\end{equation} 
Solving for $\tau$ yields the simple relation: $\tau = 2 L R_\text{out}^2 / [R_\text{in} (L^2 + R_\text{out}^2) ] $. Then, $\gamma_\text{in} = \tau \gamma_\text{out}$.

\subsection{Determination of $\ell_\parallel$}\label{}
\begin{figure}
\begin{center}
\includegraphics[width=\textwidth]{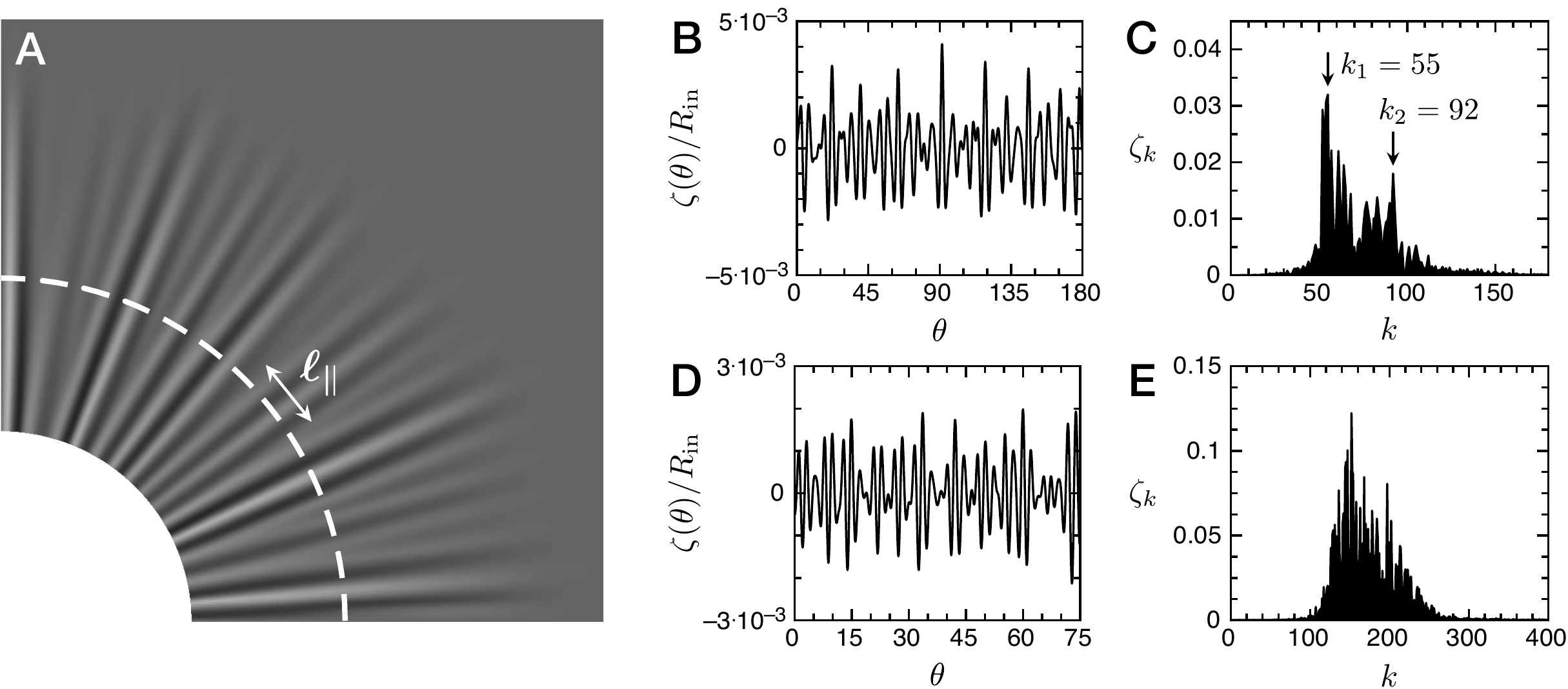}
\end{center}
\caption{
(A) Wrinkle pattern from simulations, showing strong amplitude modulation (${Bo} = 5$, $\tau=5.2$). 
Dashed line: $r/R_\text{in} = 1.8$. 
(B) Height function versus $\theta$ along a circle of radius $r = 1.8 R_\text{in}$ from the simulation in panel A. 
(C) Fourier spectrum, $\zeta_k(k)$, for the signal in panel B. 
The spectrum shows two dominant peaks. 
(D) Height function versus $\theta$ along a circle of radius $r = 1.54 R_\text{in}$ from the simulation shown in Fig.~2d in the main text (${Bo} = 32$, $\tau=5$). 
(E) Fourier spectrum for the signal in panel D. 
In this case, there are no two dominant peaks, although values of $\lambda$ and $\ell_\parallel$ can still be extracted from the image. 
}
\label{fig:SI1}
\end{figure}

In the main text, we report values of $\ell_\parallel$ that are extracted from the grayscale images of the sheet, in both experiments and simulations. 
Figure~\ref{fig:SI1} shows a typical example from simulation, where the shade corresponds to the out-of-plane displacement, with the medium tone in the upper-right corresponding to zero deflection. 
The wrinkle pattern may be decomposed into defect-free and defect-rich zones, which are alternating angular sectors of various sizes. 
The defect-rich zones are the amplitude-suppressed regions (i.e., regions with weaker contrast) where new wrinkles appear. 
The defect-free zones are characterized by nearly (but not perfectly) parallel wrinkles that have larger amplitude, hence, stronger contrast. 
Some wrinkles in these regions extend all the way from $r=R_\text{in}$ to $r=L$ without interruption. 
At a given radius, $\ell_\parallel$ is taken as the average width of the defect-free regions. 
In practice, several measurements at different locations are averaged together, and the error bar is taken to be the standard deviation of these values. 
%Some judgement is required to determine the edges that demarcate the defect-free and defect-rich regions. 
%Nevertheless, a characteristic size may be deduced from the images. 
The wrinkle wavelength serves as a natural local ``meter-stick''; we thus measure $\ell_\parallel$ in units of the observed $\lambda$ at that location. %; hence we report $\ell_\parallel/\lambda$ on the vertical axis of Fig.~4e in the main text. 

In some cases, one may also extract $\ell_\parallel$ from an analysis of the Fourier spectrum of the height function at fixed radius, $\zeta(\theta)$. 
We consider the simplest possible amplitude-modulated signal, 
\begin{equation}
\zeta(\theta) = \cos(m \theta) \cos (n \theta), 
\end{equation}
(corresponding to Eq.~\ref{eq:pure-states}), 
%Eq.~15 of the main text) 
where $m = 2\pi r/\lambda$ and $n=\pi r/\ell_\parallel$, as pictured in the schematic in Fig.~3b in the main text. 
(Note the absence of a factor of $2$ in the relation between $n$ and $\ell_\parallel$, since the size of a smectic region corresponds to a half-wavelength of the modulation envelope, see Fig. 3 of the main text.) 
This signal may be decomposed into two equal-amplitude components, $\zeta(\theta) = (1/2)\cos[(m+n)\theta] + (1/2)\cos[(m-n)\theta]$. 
Therefore, a Fourier spectrum that has two strong peaks at $k_1$ and $k_2$ is suggestive of a high-frequency component (i.e., a wrinkle number) of wavenumber $m = (k_1+k_2)/2$, with an amplitude that is modulated at a wavenumber $n = (k_2-k_1)/2$. 
We thus obtain:
\begin{equation}\label{eq:Fourier}
\frac{\ell_\parallel}{\lambda} = \frac{k_1 + k_2}{2(k_2 - k_1)}. 
\end{equation}
Figure~\ref{fig:SI1}c shows the Fourier spectrum for the image in Fig.~\ref{fig:SI1}a. 
The data show a number of spikes of various strength, but the overall trend (i.e., averaging over the noise) is perhaps best described as two strong peaks at $k_1=55$ and $k_2=92$. 
(These exact values are the locations of the maxima of $\zeta_k$ in the domains $k < 70$ and $k > 70$, respectively.) 
Plugging into Eq.~\ref{eq:Fourier} yields $\ell_\parallel/\lambda=2.0$, in agreement with the value obtained visually from the image (double-sided arrow in Fig.~\ref{fig:SI1}a).

Other images do not yield to such an analysis. 
Figs.~\ref{fig:SI1}d,e show the height function and spectrum corresponding to Fig.~2d in the main text, at $r/R_\text{in} = 1.54$. 
Here the spectrum is broad, and it is difficult to distinguish two dominant peaks from the noise, or for that matter, to determine that the number of significant peaks in the spectrum is exactly $2$. 
What we can say is that the gross shape of the spectrum is consistent with being centered around $k\approx 160$, in agreement with the wrinkle number measured at that radius using the autocorrelation method described in Section~6B above ($m=172$). 
Despite this difficulty in obtaining a value of $\ell_\parallel$ from the Fourier spectrum, an amplitude-modulation lengthscale can still be extracted from direct inspection of the image by identifying regions of strong, nearly-parallel wrinkles.

%Figure~\ref{fig:SI1}c shows the Fourier spectrum for the image in Fig.~\ref{fig:SI1}a; it shows strong peaks at $k_1=55$ and $k_2=92$. Plugging into Eq.~\ref{eq:Fourier} yields $\ell_\parallel/\lambda=2.0$, in agreement with the value obtained visually from the image (double-sided arrow in Fig.~\ref{fig:SI1}a). 

%Other images do not yield to such an analysis. Figs.~\ref{fig:SI1}d,e show the height function and spectrum corresponding to Fig.~2d in the main text. Here the spectrum is broad and does not show two dominant peaks. Nevertheless, a value of $\ell_\parallel$ can still be extracted from the image as described above. 

%%% Each figure should be on its own page

\clearpage

%\dataset{SI-Dataset-Fig1.xslx}{Height profiles in Figs.~S1b,d and Fourier spectra in Figs.~S1c,e.}
%\dataset{SI-Dataset-Fig2.xslx}{Wrinkle number data in Figs.~2e-h and phase diagram data in Fig.~2i. }
%\dataset{SI-Dataset-Fig3.xslx}{Wrinkle profiles in Figs.~4a,4c, $\ell_\parallel/\lambda$ data in Figs.~4b,4d,4e, and wrinkle number data in the Fig.~4e inset. }

%%% Add this line AFTER all your figures and tables

%\movie{Type caption for the movie here.}

%\movie{Type caption for the other movie here. Adding longer text to show what happens, to decide on alignment and/or indentations.}

%\movie{A third movie, just for kicks.}

%\dataset{dataset_one.txt}{Type or paste caption here.}

%\dataset{dataset_two.txt}{Type or paste caption here. Adding longer text to show what happens, to decide on alignment and/or indentations for multi-line or paragraph captions.}

%\bibliographystyle{pnas} 
%\bibliography{Gauss-Euler-supp}  

%\end{article}

\end{document}